\documentclass[a4paper,11pt]{article}
\usepackage{jheppub} 
\usepackage[normalem]{ulem}
\usepackage{slashed}
\usepackage{bigints}
\usepackage{braket}
\usepackage{comment}
\usepackage{bm}		    

\def\be{\begin{equation}}
\def\ee{\end{equation}}
\def\figs/B{B}
\def\bea{\begin{eqnarray}}
\def\eea{\end{eqnarray}}
\def\bg{\begin{eqnarray}}
\def\nd{\end{eqnarray}}

\newcommand{\xvec}{{\bm x}}
\newcommand{\kvec}{{\bm k}}

\newcommand{\pl}{{\mbox{\tiny P}}}

\newcommand{\rh}{{\mbox{\tiny RH}}}
\newcommand{\MPl}{M_\mathrm{Pl}}
\newcommand{\early}{{\mbox{\tiny early}}}
\newcommand{\late}{{\mbox{\tiny late}}}

\title{\boldmath Gravitational Production of Completely Dark Photons with Nonminimal Couplings to Gravity}

\author[a]{Christian Capanelli,}
\author[b]{Leah Jenks,}
\author[b,c]{Edward W. Kolb,}
\author[d]{Evan McDonough}

\affiliation[a]{Department of Physics \& Trottier Space Institute,\\ 
McGill University, Montr\'eal, QC H3A 2T8, Canada.}
\affiliation[b]{Kavli Institute for Cosmological Physics, \\
The University of Chicago,  5640 South Ellis Avenue, Chicago, IL 60637, U.S.A.}
\affiliation[c]{Enrico Fermi Institute,\\
The University of Chicago,  5640 South Ellis Avenue, Chicago, IL 60637, U.S.A.}
\affiliation[d]{Department of Physics,\\ University of Winnipeg, Winnipeg MB, R3B 2E9, Canada}

\emailAdd{christian.capanelli@mail.mcgill.ca}
\emailAdd{ljenks@uchicago.edu}
\emailAdd{Rocky.Kolb@uchicago.edu}
\emailAdd{e.mcdonough@uwinnipeg.ca}

\abstract{Dark photons are a theorized massive spin-1 particle which can be produced via various mechanisms, including cosmological gravitational particle production (GPP) in the early universe. In this work, we extend previous results for GPP of dark photons to include nonminimal couplings to gravity. We find that nonminimal couplings can induce a ghost instability or lead to runaway particle production at high momentum and discuss the constraints on the parameter space such that the theory is free of instabilities. Within the instability-free regime we numerically calculate the particle production and find that the inclusion of nonminimal couplings can lead to an enhancement of the particle number. As a result, GPP of nonminimally coupled dark photons can open the parameter space for production of a cosmological relevant relic density (constituting all or part of the dark matter) as compared to the minimally-coupled theory. 
These results are independent of the choice of inflation model, which we demonstrate by repeating the analysis for a class of rapid-turn multi-field inflation models. }

\begin{document}
\maketitle
\flushbottom

\section{Introduction}
\label{sec:intro}
Dark matter (DM) comprises approximately a quarter of the present energy density of our universe. However, despite compelling observational evidence for its existence, it has yet to be directly detected, and its exact nature is still unknown. The available parameter space for the properties of dark matter and the mechanisms by which it might be produced are vast. There are myriad dark matter candidates which can be produced via interactions with the standard model (SM) in the early universe. Two well-studied methods to produce dark matter are freeze-out, in which the dark matter begins in thermal equilibrium with SM particles then `freezes-out' to some relic abundance due to the expansion of the universe, and `freeze-in' in which the dark matter does not begin in thermal equilibrium, but rather is produced due to small interactions with standard model particles. Both of these processes require the dark matter to interact with the standard model. Given that one of the few facts known about dark matter is that it must interact only very weakly with the standard model, if at all, it is attractive to consider alternative dark matter production mechanisms which do not require any couplings to standard model particles, e.g., which are `completely dark.' 

One such mechanism is cosmological gravitational particle production (GPP) \cite{Parker:1969au, Parker:1971pt, Ford:2021syk,Chung:1998zb,Chung:1998ua} in which particles are produced during inflation due to the rapid expansion of the universe. Particles produced via GPP are generally thought to be in the supermassive range (i.e., $m\gtrsim H_e$ where $H_e$ is the Hubble parameter at the end of inflation), but can also have masses in the ultralight regime \cite{Kolb:2020fwh,Graham:2015rva,Ahmed:2020fhc}. Previous work has shown that GPP is a viable production mechanism for particles with a wide variety of masses and spins \cite{Chung:1998zb, Graham:2015rva, Ema:2015dka, Ema:2016hlw, Ema:2019yrd, Kolb:2020fwh,Ahmed:2020fhc, Alexander:2020gmv,Kolb:2021xfn,Kolb:2021nob,Kolb:2023dzp, Capanelli:2023uwv,Kaneta:2023uwi,Maleknejad:2022gyf} and in both single field and multifield inflation models \cite{Kolb:2022eyn}. In particular, Refs.\ \cite{Graham:2015rva, Kolb:2020fwh} have shown that spin-1, or `dark photon' dark matter can be gravitationally produced to yield the correct DM relic density. Dark photons have been considered as a portal from the standard model to a dark sector, but can also be considered a DM candidate in their own right, see e.g., \cite{Redondo_2009, Nelson:2011sf,Long:2019lwl,Arias:2012az,Graham:2015rva, Kolb:2020fwh,Co:2018lka,Dror:2018pdh,Bastero-Gil:2018uel, Bastero-Gil:2021wsf, Agrawal:2018vin, Krnjaic:2022wor, Pospelov:2008jk, Fradette_2014,Jaeckel_2013,Nakayama:2019rhg,East:2022rsi,Cyncynates:2023zwj,Zhang:2023fhs,Kitajima:2023fun, Bhattacharyya:2023kbh}. 

Previous work on GPP of dark photon dark matter has largely considered particles which are minimally coupled to gravity \cite{Graham:2015rva, Ema:2019yrd, Kolb:2020fwh, Ahmed:2020fhc}. However, a theory of a massive spin-1 field can also contain nonminimal couplings to gravity; general principles of effectice field theory (EFT) demand their inclusion. 
These additional couplings are an important factor in fully understanding gravitational production of spin-1 particles and dark photon dark matter. Recent works have begun to discuss these possibilities \cite{Alonso-Alvarez:2019ixv, Capanelli:2023uwv, Ozsoy:2023gnl, Cembranos:2023qph}, however a full survey of the viable parameter space while also taking into account instabilities in the theory is yet to appear in the literature. 

In this paper, we extend previous work to consider the gravitational production of dark photon dark matter with nonminimal couplings to gravity. We carefully consider the viable parameter space of the theory which is both non-ghostly and does not lead to runaway particle production due to a tachyonic gradient instability (as discussed in \cite{Capanelli:2024pzd}) and show how the addition of nonminimal couplings in the allowed region can lead to an enhancement of particle production, particularly at low masses, while also avoiding modes which propagate superluminally. We find the parameter space to obtain the corrent present-day dark matter relic density and find that the addition of nonminimal couplings allows for gravitational production of dark matter with the correct relic abundance, extending the possible range of parameter space found for the minimal theory. We further discuss the GPP of dark photons in a broader class of rapid-turn multi-field inflation models in the minimally coupled theory.

The structure of the paper is as follows: After a brief review of GPP in Sec.\ \ref{sec:GPPABC}, in Sec.\ \ref{sec:proca} we discuss the theory of non-minimally coupled spin-1 fields, then in Sec.\ \ref{sec:instabilities} we discuss the conditions on the parameter space such that the theory remains ghost-free and also avoids catastrophic runaway production. We then show numerical results in Sec.\ \ref{sec:results}, and discuss the present-day dark matter relic density and allowed parameter space to be a viable dark matter candidate. Finally, we conclude with a discussion in Sec.\ \ref{sec:conclusions}. Lastly, we show explicit GPP results for dark photons in rapid-turn multi-field inflation in Appendix~\ref{sec:GPP-multi} and for early reheating in Appendix~\ref{sec:early}.

Throughout this work the following conventions are employed: we use a mostly minus metric signature and natural units $c = \hbar = 1$ unless explicitly stated otherwise. Greek letters $\mu, \nu, ...$ indicate a sum over all spacetime indices while Latin letters $i,j,k$ denote a sum over spatial indices. 

\section{ABCs of Gravitational Particle Production}
\label{sec:GPPABC}
Here we give a brief overview of the mechanics of gravitational particle production. For further details, see \cite{Kolb:2023ydq} and references therein. As discussed previously, gravitational particle production exploits the fact that particles can be created by the expansion of the universe.  This is due to the fact that the initial (early) and final (late) creation and annihilation operators are not the same, but can be related by a Bogliubov transformation with the coefficients $\alpha_k$ and $\beta_k$. 

Solutions to the wave equation can be expressed in terms of mode functions $\chi_\kvec(\eta)$ as 
\begin{align}
\phi(\eta,\xvec) = \int{\frac{d^3\kvec}{(2\pi)^3}} \left[\hat{a}_\kvec\chi_\kvec(\eta) e^{i\kvec\cdot\xvec} + \hat{a}_\kvec^\dagger \chi^*_\kvec(\eta)e^{-i\kvec\cdot\xvec} \right] \ .
\end{align}
If one defines the initial vacuum state with early-time creation and annihilation operators $\hat{a}_\kvec^\dagger$ and $\hat{a}_\kvec$ as 
\be 
\hat{a}_\kvec^\early  \ket{0^\early} = 0\ket{0^\early }, 
\ee 
which is related to the late-time operators as 
\begin{align}
\hat{a}_\kvec^\early = \alpha_\kvec^* \hat{a}_\kvec^\late - \beta_\kvec^* \hat{a}_{-\kvec}^{\late\dagger},
\end{align}
where $\alpha_k$ and $\beta_k$ are the Bogoliubov coefficients. The value of $\beta_k$ can be found from the mode equations such that 
\be 
\label{eq:betak}
|\beta_{k}|^2 = \lim_{\eta \rightarrow \infty} \left[\frac{\omega_k}{2}|\chi_k|^2 + \frac{1}{2\omega_k}|\partial_\eta\chi_k|^2 - \frac{1}{2}\right], 
\ee 
where $\omega_k$ is the frequency of the mode. (Since the background spacetime will be taken to be homogeneous and isotropic, $\beta$ depends only on $k=|\kvec|$.) Then, we can construct the spectrum of particles produced in terms of the $\beta_k$ and the comoving wavenumber as \cite{Kolb:2020fwh}
\begin{equation}
n_k =\frac{k^3}{2\pi^2}|\beta_k|^2. 
\end{equation}
The comoving number density is then given by 
\be 
na^3 = \int \frac{dk}{k} n_k.
\ee 
From the comoving number density, one can determine the relic density, as we will explicitly see in Sec.~\ref{sec:relic}. Generally, one takes Bunch-Davies initial conditions to numerically solve for the particle production. The early time limit $\eta \rightarrow -\infty$ corresponds to the limit where $a\rightarrow 0$ and $a^2R\rightarrow$0, which implies that $\omega_k^2\rightarrow k^2$. In this limit, the mode functions are are deep within the Hubble radius. The Bunch-Davies initial conditions then correspond to mode equations that are in a Minkowski spacetime, that is
\begin{equation}
    \lim _{k\eta \rightarrow - \infty} \chi_k(\eta)=\frac{1}{\sqrt{2 k}} e^{-i k \eta}.
\end{equation}

\section{Spin-1 Field with Nonminimal Couplings }
\label{sec:proca}
In this section we review details of de Broglie-Proca theory for a massive spin-1 field with nonminimal couplings to gravity \cite{deBroglie:1922zz, deBroglie:1924ldk, Proca:1936fbw}.  On a generic spacetime background with metric $g_{\mu\nu}$, the de Broglie-Proca action for a massive spin-1 (dark photon) field can be written as \cite{Kolb:2020fwh}
\be
S =  \int d^4x \sqrt{-g} \left(-\frac{1}{4}F^{\mu\nu}F_{\mu\nu} + \frac{1}{2}m^2 g^{\mu\nu}A_\mu A_\nu - \frac{1}{2}\xi_1 R g^{\mu\nu}A_\mu A_\nu - \frac{1}{2} \xi_2 R^{\mu\nu}A_\mu A_\nu\right), 
\label{eq:Sproca}
\ee 
where $A_\mu$ is the dark photon, $R$ the Ricci scalar, $R_{\mu\nu}$ the Ricci tensor and $\xi_1$ and $\xi_2$ are dimensionless coupling constants. The dark photon field strength, $F_{\mu\nu}$, is defined as $F_{\mu\nu} = \nabla_\mu A_\nu - \nabla_\nu A_\mu$. 

Notice that the two interaction terms we have constructed are the only dimension-four operators which can appear with the vector field coupled to curvature. They are consistent with the symmetries of the Proca theory and Einstein gravity, and thus, the effective field theory approach would be to include these terms with coupling constants $\xi_1$ and $\xi_2$ to be fixed by experiment. Moreover, as discussed in \cite{Capanelli:2024pzd}, these terms are expected to be generated by loops even if they are set to zero at tree-level. This is analogous to the renormalization of scalar fields in curved space \cite{Parker:2009uva,Birrell:1982ix,Bunch:1980bs}.

In principle, additional couplings can appear, for example to the Riemann tensor, but will involve higher dimensional operators. Thus, we consider only the two coupling terms as written above. To recover the minimally coupled spin-1 scenario, one can simply set $\xi_1 = \xi_2 = 0$. As we will see below, the two terms proportional to $\{\xi_1, \xi_2\}$ will induce a time-dependent effective mass for the dark photon.   

Let us now specify these general considerations to a cosmological background. We now take $g_{\mu\nu}$ to be the Friedmann-Lemaitre-Robertson-Walker (FLRW) metric, $g_{\mu\nu} = a^2(\eta)\text{diag}(1, -1, -1, -1)$, where $a(\eta)$ is the scale factor as a function of conformal time. Following \cite{Kolb:2020fwh}, we can explicitly see the effects of the non-minimal couplings $\xi_1$ and $\xi_2$ by decomposing the action, Eq.~\eqref{eq:Sproca}, into components $A_0$ and $A_i$:
\begin{equation}
\begin{aligned}
S\left[A_\mu(t, \boldsymbol{x})\right]= & \int d^4 x\left[\frac{1}{2} a\left(\partial_0 A_i-\partial_i A_0\right)^2-\frac{1}{4} a^{-1}\left(\partial_i A_j-\partial_j A_i\right)^2\right. \\
& \left.+\frac{1}{2} a^3 m_{t}^2 A_0^2-\frac{1}{2} a m_{x}^2 A_i^2\right],
\end{aligned}
\end{equation}
where we have defined two effective masses, $m_t$ and $m_x$, as
\begin{align}
    m^2_{t} = m^2 - \xi_1R - \frac{1}{2}\xi_2 R - 3 \xi_2 H^2 \label{eq:mefft}\\
    m^2_{x} = m^2 - \xi_1R - \frac{1}{6}\xi_2R + \xi_2H^2. \label{eq:meffx}
\end{align}
In this expansion, we can see the explicit role of $\xi_1$ and $\xi_2$ as contributions to the effective masses of $A_0$ and $A_i$, and furthermore that $m_t$ and $m_x$ will now be \textit{time-dependent} functions given that $R$ and $H$ evolve throughout the inflationary period.

Let us further expand the action in terms of mode functions
\be 
A^{\mu}(t, \xvec) = \int \frac{d^3 \kvec}{(2\pi)^3} A_{\kvec}^\mu e^{i \kvec\cdot\xvec}
\ee 
to obtain 
\begin{equation}
\begin{aligned}
S\left[A_\mu(t, \boldsymbol{x})\right]= & \int d t \int \frac{d^3 \boldsymbol{k}}{(2 \pi)^3}\left[\frac{i}{2} a \boldsymbol{k}_i A_0^*\left(\partial_0 A_i\right)-\frac{i}{2} a \boldsymbol{k}_i\left(\partial_0 A_i^*\right) A_0+\frac{1}{2} a\left(k^2+a^2 m^2_{t}\right)\left|A_0\right|^2\right. \\
& \left.-\frac{1}{4} a^{-1}\left|\boldsymbol{k}_i A_j-\boldsymbol{k}_j A_i\right|^2+\frac{1}{2} a\left|\partial_0 A_i\right|^2-\frac{1}{2} a m^2_{x}\left|A_i\right|^2\right],
\end{aligned}
\end{equation}
where we have integrated over $\kvec'$ and $\xvec$, leaving integration over only $\kvec$, and have suppressed the $\kvec$ subscript on $A_0$ and $A_i$ for notational simplicity. 

In order to solve for $A_0$, we rewrite the action as
\begin{equation}
\begin{aligned}
S\left[A_\mu(t, \boldsymbol{x})\right] & =\int d t \int \frac{d^3 \boldsymbol{k}}{\left(2 \pi^3\right)}\left[\frac{1}{2} a\left(k^2+a^2 m^2_{t}\right)\left|A_0+i \frac{\boldsymbol{k}_i\left(\partial_0 A_i\right)}{k^2+a^2 m^2_{t}}\right|^2\right. \\
& \left.-\frac{1}{2} a \frac{\left|\boldsymbol{k}_i\left(\partial_0 A_i\right)\right|^2}{k^2+a^2 m^2_{t}}-\frac{1}{4} a^{-1}\left|\boldsymbol{k}_i A_j-\boldsymbol{k}_j A_i\right|^2+\frac{1}{2} a\left|\partial_0 A_i\right|^2-\frac{1}{2} a^2 m^2_{x}\left|A_i\right|^2\right].
\end{aligned}
\end{equation}
In this form, it is clear that the temporal component, $A_0$, is non-dynamical, and can be solved for as:
\be
A_0=-i \frac{\boldsymbol{k}_i\left(\partial_0 A_i\right)}{k^2+a^2 m^2_{t}}.
\ee
Integrating out $A_0$, the action becomes
\begin{equation}
\begin{aligned}
S\left[A_\mu(t, \boldsymbol{x})\right] & =\int d t \int \frac{d^3 \boldsymbol{k}}{\left(2 \pi^3\right)}\left[\frac{1}{2} a\left(\partial_0 A_i^*\right)\left(\delta_{i j}-\frac{\boldsymbol{k}_i \boldsymbol{k}_j}{k^2+a^2 m^2_{t}}\right)\left(\partial_0 A_j\right)\right. \\
& \left.-\frac{1}{2} a^{-1} A_i^*\left[\left(k^2+a^2 m^2_{x}\right) \delta_{i j}-\boldsymbol{k}_i \boldsymbol{k}_j\right] A_j\right] ,
\end{aligned}
\end{equation}
By introducing an orthonormal set of transverse and longitudinal mode functions $A_{\boldsymbol{k}}^T\left(x^0\right)$ and $A_{\boldsymbol{k}}^L\left(x^0\right)$, the action separates into two pieces:
\begin{equation}
\begin{aligned}
S^T & =\sum_{b=1,2} \int d \eta \int \frac{d^3 \boldsymbol{k}}{(2 \pi)^3}\left[\frac{1}{2}\left|\partial_\eta A_{\boldsymbol{k}}^{T_b}\right|^2-\frac{1}{2}\left(k^2+a^2 m^2_{x}\right)\left|A_{\boldsymbol{k}}^{T_b}\right|\right] \\
S^L & =\int d \eta \int \frac{d^3 \boldsymbol{k}}{(2 \pi)^3}\left[\frac{1}{2} \frac{a^2 m^2_{t}}{k^2+a^2 m^2_{t}}\left|\partial_\eta A_{\boldsymbol{k}}^L\right|^2-\frac{1}{2} a^2 m^2_{x}\left|A_{\boldsymbol{k}}^L\right|^2\right], 
\end{aligned}
\end{equation}
where we have now expressed $S_L$ and $S_T$ in terms of the conformal time, $\eta$. In order for the kinetic term of $S_L$ to be canonically normalized, we perform a field redefinition of the longitudinal mode such that
\be
A_{\boldsymbol{k}}^L(\eta)=\kappa(\eta) \chi_{\boldsymbol{k}}^L(\eta), 
\ee 
where 
\be 
\kappa^2(\eta)={\frac{k^2+a^2 m^2_{t}}{a^2 m^2_{t}}}.
\label{eq:kappasq}
\ee
Finally, we are left with two decoupled actions for the transverse and longitudinal modes: 
\begin{equation}
\begin{aligned}
S^T & =\sum_{b=1,2} \int d \eta \int \frac{d^3 \boldsymbol{k}}{(2 \pi)^3}\left(\frac{1}{2}\left|\partial_\eta A_{\boldsymbol{k}}^{T_b}\right|^2-\frac{1}{2} \omega_T^2\left|A_{\boldsymbol{k}}^{T_b}\right|^2\right) \\
S^L & =\int d \eta \int \frac{d^3 \boldsymbol{k}}{(2 \pi)^3}\left(\frac{1}{2}\left|\partial_\eta \chi\right|^2-\frac{1}{2} \omega_L^2|\chi|^2\right).
\label{eq:SLfinal}
\end{aligned}
\end{equation}
The transverse and longitudinal frequencies, $\omega_T$ and $\omega_L$, respectively, are defined as:
\begin{align}
    \omega_T^2(\eta) &= k^2 + a^2 m^2_{x} \label{eq:omegaT}\\
    \omega_L^2(\eta) &= k^2 \frac{m^2_{x}}{m^2_{t}} + a^2 m^2_{x} + \partial_\eta \left(\frac{\partial_\eta \kappa}{\kappa}\right) - \left(\frac{\partial_\eta \kappa}{\kappa}\right)^2. \label{eq:omegaL}
\end{align}
In the minimally coupled theory, the frequencies reduce to the explicit forms found in Ref.\ \cite{Kolb:2020fwh}, but become substantially more complicated with $\{\xi_1, \xi_2\} \neq 0$. We have:
\begin{align}
    \omega_T^2 &= k^2 + a^2 m^2_{ x} \\
    \omega_L^2 &= k^2 \frac{m^2_{ x}}{m^2_{t}} + a^2 m^2_{x} + \frac{3k^2a^4 m^2_{t}H^2}{(k^2 + a^2m^2_{t})^2} + \frac{k^2a^2R}{6(k^2 + a^2m^2_{t})} \nonumber \\
    &+ \frac{H a k^2 m_{t}^{2'}}{m^2_{t}}\frac{(-k^2 + 2 a^2 m^2_{t})}{(k^2 + a^2m^2_{t})^2} + \frac{k^2(m_{t}^{2'})^2}{4 (m^2_{t})^2}\frac{(k^2 + 4 a^2 m^2_{t})}{(k^2 + a^2m^2_{t})^2} - \frac{k^2 m_{t}^{2''}}{2m^2_{t}(k^2 + a^2 m^2_{t})}.
  \label{eq:omegaLfull}
\end{align}
where the derivatives of $m_{t}$ are given explicitly by 
\begin{align}
    m^{2'}_{t} &= -6\xi_2 H H' - \xi_1 R' - \frac{1}{2}\xi_2 R', \\
    m^{2''}_{t} &=-6\xi_2 H'^2 -6\xi_2 H H'' - \xi_1 R'' - \frac{1}{2}\xi_2R''.
\end{align}
Taking $\{\xi_1, \xi_2\} = 0$, we are left with $m_x = m_t = m$ and $ m^{2'}_{t} =  m^{2''}_{t} =0$, recovering the result in the minimal theory. We can thus see that the effects of the non-minimal modifications are controlled by the choice of the coupling parameters, $\{\xi_1, \xi_2\}$.

A signature feature of this model, as can be appreciated from the $k^2$ term in $\omega_ L ^2$, is the time-dependent sound speed of the longitudinal mode,
\begin{equation}
    c_s ^2 \equiv \frac{m_x ^2}{m_t^2} =  \frac{1 - \frac{\xi_1 + \frac{1}{6}\xi_2}{m^2} R + \frac{\xi_2}{m^2} H^2}{1 - \frac{\xi_1+\frac{1}{2}\xi_2}{m^2} R - 3 \frac{\xi_2}{m^2} H^2 }
    \label{eq:cs2}
\end{equation}
From this one may appreciate that the sound speed depends only on the mass and couplings via the combination $\xi_{1,2}/m^2$. 

In the spirit of effective field theory, these couplings are free parameters to be fixed by comparison to data. However, as we will see in the next section, if one desires a healthy theory which does not propagate ghosts and tachyonic instabilities, more restrictions apply.

\section{Instabilities of the Theory}
\label{sec:instabilities}

We now turn to the possible instabilities in the Proca theory with nonminimal couplings\footnote{An abbreviated discussion of instabilities in Proca with nonminimal couplings can be found in Ref.\ \cite{Capanelli:2024pzd}.}. As we will see, a key distinction between the minimally and nonminimally coupled theories is that with the inclusion of nonminimal couplings, there are several potential instabilities. This can be seen explicitly from the forms of the effective masses, Eqs.~\eqref{eq:mefft} and \eqref{eq:meffx}, which are not necessarily positive definite for all values of $R$ and $H$ and in fact oscillate through zero for particular combinations of $\{m/H_e,\xi_1,\xi_2\}$.\footnote{The subscript ``e'' indicates the value of the variable at the end of inflation.} As a result, there are three potential instabilities: 
\begin{enumerate}
    \item {\bf Ghost:} The theory will propagate ghost modes due to an overall minus sign in the kinetic term of the longitudinal mode. 
    \item {\bf Gradient:} At high momentum, a negative value of $m_{x}^2$ will lead to a runaway particle production \cite{capanelli2024runaway}.
    \item {\bf Superluminal:} For parameters which are both ghost-free and runaway-free, it is possible for modes to have sound speed $c_s > 1$ and propagate superluminally.
\end{enumerate}  
We would like to understand the viable parameter space in which dark photons can be gravitationally produced without any of these instabilities, i.e., ghost-free, non-runaway, and subluminal. We will refer to such a regions as a ``safe'' region. Let us consider each of these scenarios in turn.

\subsection{Ghost Instabilities}
\label{sec:ghosts}
First let us explore the scenario which leads to a ghost instability in the theory. The presence of a ghost is indicated by the kinetic term of the action Eq.~\eqref{eq:SLfinal} having the wrong sign. This is controlled by the parameter $\kappa$, Eq.~\eqref{eq:kappasq}, and therefore the effective mass $m_t^2$. Clearly, for positive values of $m_t^2$ there will be no issues as $\kappa^2$ will remain positive throughout the evolution from inflation into radiation and/or matter domination. Negative values of $\kappa^2$, however, will be problematic. In particular, if $m_t^2 <0$ and $k^2 > a^2 m_t^2$, then $\kappa^2 <0$, leading to a ghost. Notice that it is indeed possible for $m_t^2$ to be negative but retain $\kappa^2 >0$ if one considers low-momentum modes with $k^2 < a^2 m_t^2$. However, we would like for the theory to remain ghost-free for arbitrarily large $k$. Therefore, the conservative requirement for a ghostless theory is to demand that  $m^2_{t}$ remains positive definite for given values of $\{\xi_1, \xi_2\}$ throughout the de Sitter phase of inflation and through the evolution into radiation and/or matter domination via reheating. 

\begin{figure}[htb!]
    \centering
    \includegraphics[width=\textwidth]{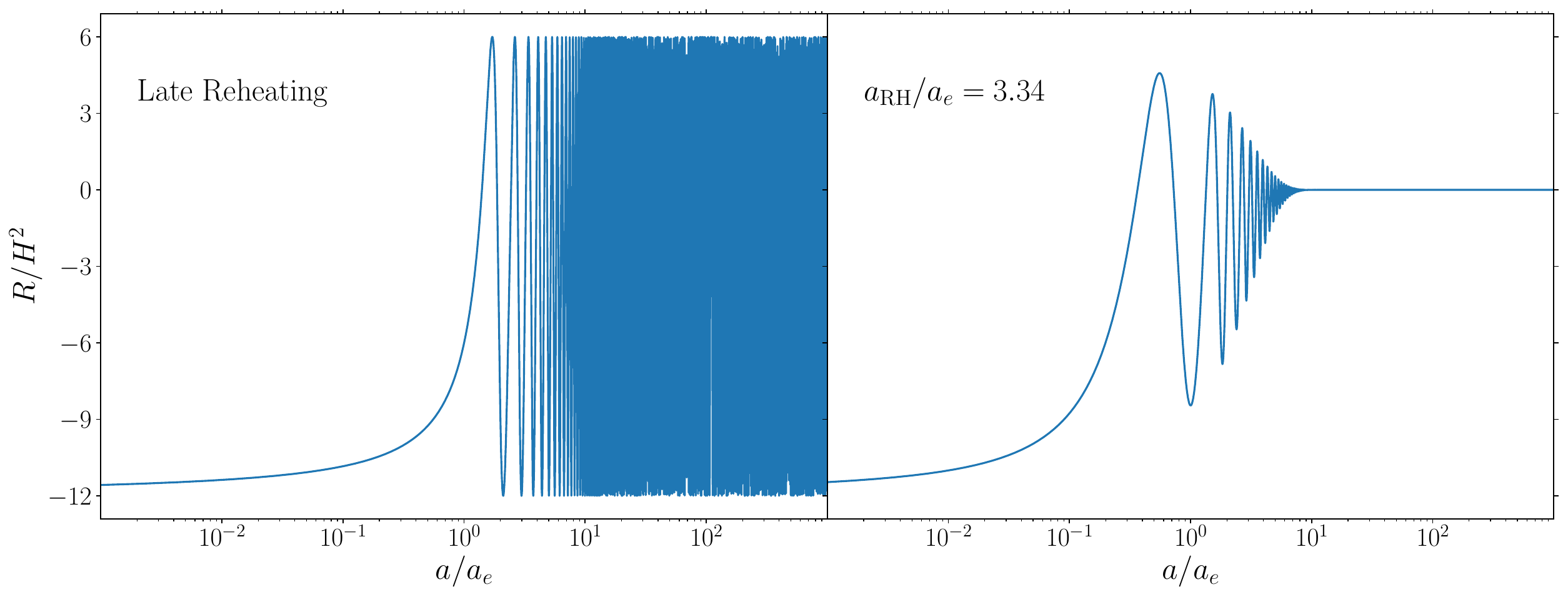}
    \caption{Evolution of $R/H^2$ for quadratic inflation in late (left) and early (right) reheating scenarios.}
    \label{fig:rbyh2}
\end{figure}

To determine where this region lies in the $\{\xi_1, \xi_2\}$ parameter space, consider first the limit in which $ m \ll H$.\footnote{It is convenient to work with dimensionless variables for the masses.  We define $\{\mu,\mu_t,\mu_x\}$ as $\{m/H_e,m_t/H_e,m_x/H_e\}$, and $\{\bar\mu,\bar\mu_t,\bar\mu_x\}$ as $\{m/H,m_t/H,m_x/H\}$. Note that $\{\mu,\mu_t,\mu_x\}$ are constant in the evolution of the background, but $\{\bar\mu,\bar\mu_t,\bar\mu_x\}$ evolve in the evolution. } Let us define the quantity $\bar{\mu}_t = m_t/H$ such that we can rewrite the expression for $m_t^2$, Eq.~\eqref{eq:mefft}, as 
\be 
\bar{\mu}_t^2(m\ll H) = -\frac{R}{H^2}\left(\xi_1 + \frac{1}{2}\xi_2\right) - 3\xi_2.
\ee 
We can see that the overall sign of $\bar{\mu}^2_t$ depends on the ratio of $R/H^2$. To see explicitly the bounds this places on $\{\xi_1, \xi_2\}$ consider a representative example of quadratic inflation such that $V_\phi \propto m^2_\phi \phi^2$, with $\phi$ the inflaton.\footnote{Though quadratic inflation model has been ruled out by Planck CMB observations  \cite{Planck:2015sxf, Planck:2018jri}, our results are largely model-independent and we therefore consider quadratic inflation as a simple, representative example. The analysis presented here also applies to allowed inflation models.} Figure~\ref{fig:rbyh2} shows the evolution of $R/H^2$ for quadratic inflation for two reheating scenarios. In the left panel, we consider `late rehating' in which reheating occurs at a late enough time that during the relevant timescales for particle production the universe evolves from an inflationary de Sitter phase into a matter-dominated phase, then reheats at a much later time. On the right, we consider an `early reheating' scenario in which there is an intermediate period of radiation domination, whose onset is controlled by the paramter $a_{\rh}/a_e$, where $a_e$ is the scale factor at the end of inflation and $a_{\rh}$ is the scale factor at equality of radiation and inflaton energy densities. Details of the early-reheating scenario are to be found in App.\ \ref{sec:early}.

In both scenarios, we see that $\text{min}(R/H^2) = -12$, corresponding to the universe in a de Sitter phase during inflation. In the late reheating scenario, $R/H^2$ oscillates to a maximum of $R/H^2 = 6$ in the matter dominated phase, corresponding to when the inflaton is at the minimum of the potential and in a momentary kination phase with $w = 1$. For the early reheating scenario, the maximum value is slightly lower than $R/H^2 = 6$ due to the intermediate radiation-dominated period. 

\begin{figure}[htb!]
\centering
    \includegraphics[width=0.9\textwidth]{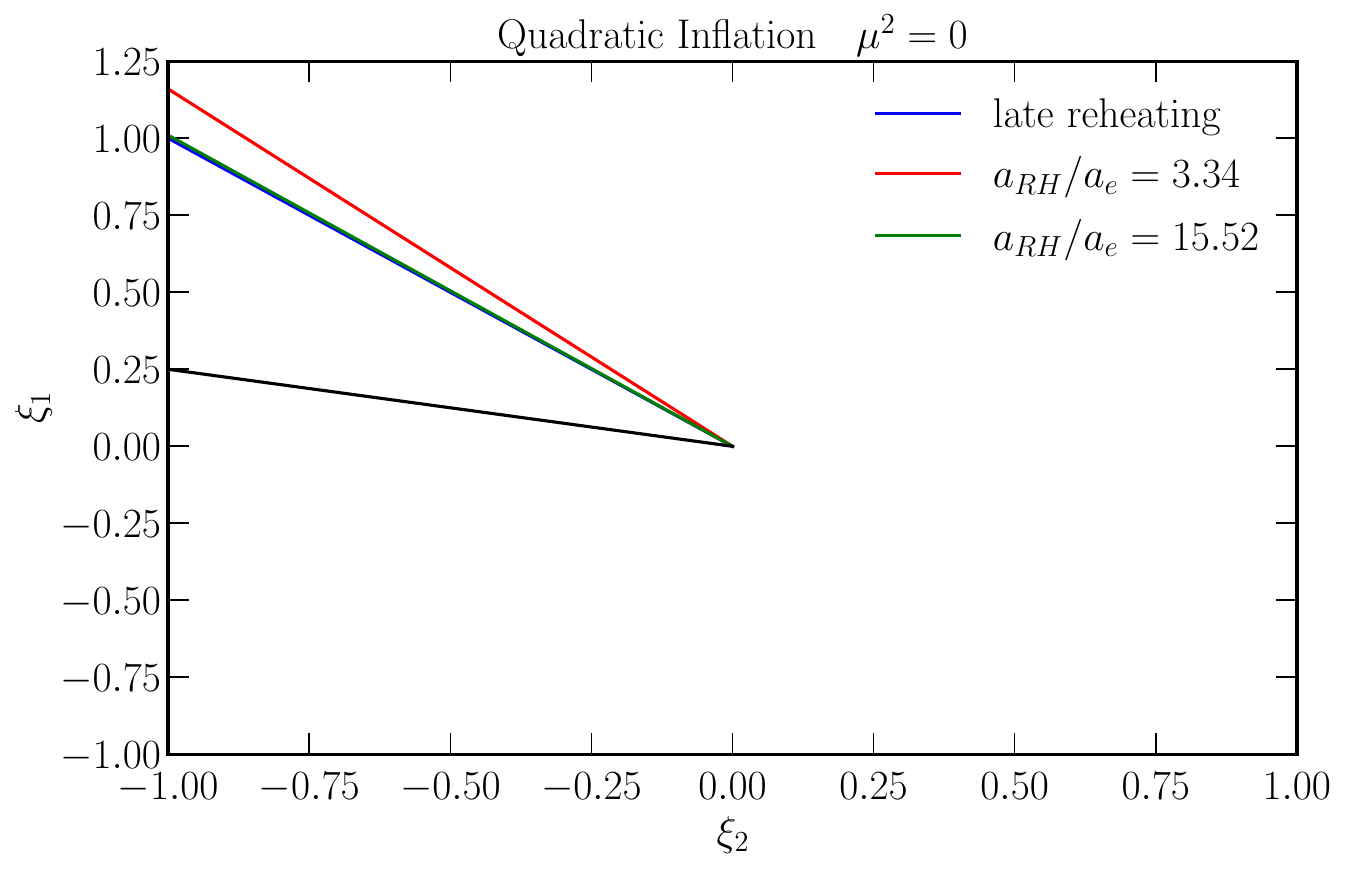}
    \caption{Ghost-free region for $m =0$ for both the late and early reheating scenarios. The upper line for $a_\mathrm{RH}/a_e=15.52$ is practically indistinguishable from the late-reheating line.}
    \label{fig:m0noghosts}
\end{figure}

We can convert these bounds into analytic constraints on $\{\xi_1, \xi_2\}$. For late reheating, requiring that $\bar{\mu}_t^2$ remain positive throughout the evolution, we find that 
\be 
-\frac{\xi_2}{4} < \xi_1  < -\xi_2, 
\ee 
which carves out a wedge in the $\{\xi_1, \xi_2\}$ plane for which the theory is non-ghostly. For early reheating, the size of the wedge increases slightly. For the choice of $a_\rh/a_e = 3.34$, we instead obtain a constraint  $-\xi_2/4 < \xi_1  \lesssim -0.866\xi_2$. Figure~\ref{fig:m0noghosts} shows the ghost-free parameter space for both of these scenarios, as well as the scenario in which $a_{\rh}/a_e = 15.52$, which is indistinguishable from the late reheating model. For most of the paper, we will focus on late reheating as a representative scenario and postpone discussion of the effects of early reheating until App.\ \ref{sec:early}.

Recall that what we have discussed above is the $m \ll H$ limit. As one increases the mass, the ghost-free region in the $\{\xi_1, \xi_2\}$ plane will also increase. We can approximate the $\bar{\mu}$-dependent condition as 
\begin{align}
\bar{\mu}^2 \gtrsim 6(\xi_1 + \xi_2) && \bar{\mu}^2 \gtrsim -12\left(\xi_1 + \frac{\xi_2}{4}\right).
\label{eq:munoghosts}
\end{align} 
Recalling that $\bar{\mu}\equiv m/H$ is itself time-dependent, since it is dependent on $H$, we require that Eq.~\eqref{eq:munoghosts} remain satisfied throughout the entire evolution. Thus, choosing $\{\xi_1, \xi_2\}$ such that Eq.~\eqref{eq:munoghosts} is satisfied for any given value of $m$ will ensure that theory remains ghostless.

\subsection{Gradient (``Sound Speed") Instability}
\label{sec:runaway}

Requiring a theory which does not propagate ghosts is a necessary, but not sufficient requirement for the theory to be stable. The ghost instability in the previous section appears when the kinetic term of the action has the incorrect sign, but there is also the possibility for a tachyonic instability in which the mass term of the action becomes negative. In this case, the modes become tachyonic, corresponding to $\omega^2 < 0$. 

Note that a transient negative $\omega^2$ is is not inherently problematic and there is a regime in which this is allowed; in fact, the oscillations of $\omega_{T,L}$ through zero lead to an enhancement of the particle production, which occurs even in the minimally coupled theory. 

To see where the tachyonic behavior becomes problematic, consider the $k \rightarrow \infty$ limit of the frequencies $\omega_{T,L}^2$, Eqs.~\eqref{eq:omegaT} and \eqref{eq:omegaL}. We can see that for the transverse mode $\lim_{k\to \infty} \omega_T^2 = k^2$ and is thus well behaved in the large-$k$ limit. On the other hand, for the longitudinal mode
\be 
\lim_{k \to \infty } \omega_L^2 = c_s ^2 k^2, 
\ee 
where $c_s^2 \equiv m_x^2 /m_t ^2$ as per Eq.~\ref{eq:cs2}. The inclusion of the nonminimal couplings induces a modification to the sound speed with $c_s = m_x/m_t \neq 1$.  We have required that $m_t^2 >0$ to remain ghost-free, but have placed no such constraints on $m_x^2 <0$. If $m_x^2<0$ while $m_t^2>0$, then $c_s^2 <0$;  referred to as a gradient instability.

\begin{figure}[htb!]
    \centering
    \includegraphics[width=\textwidth]{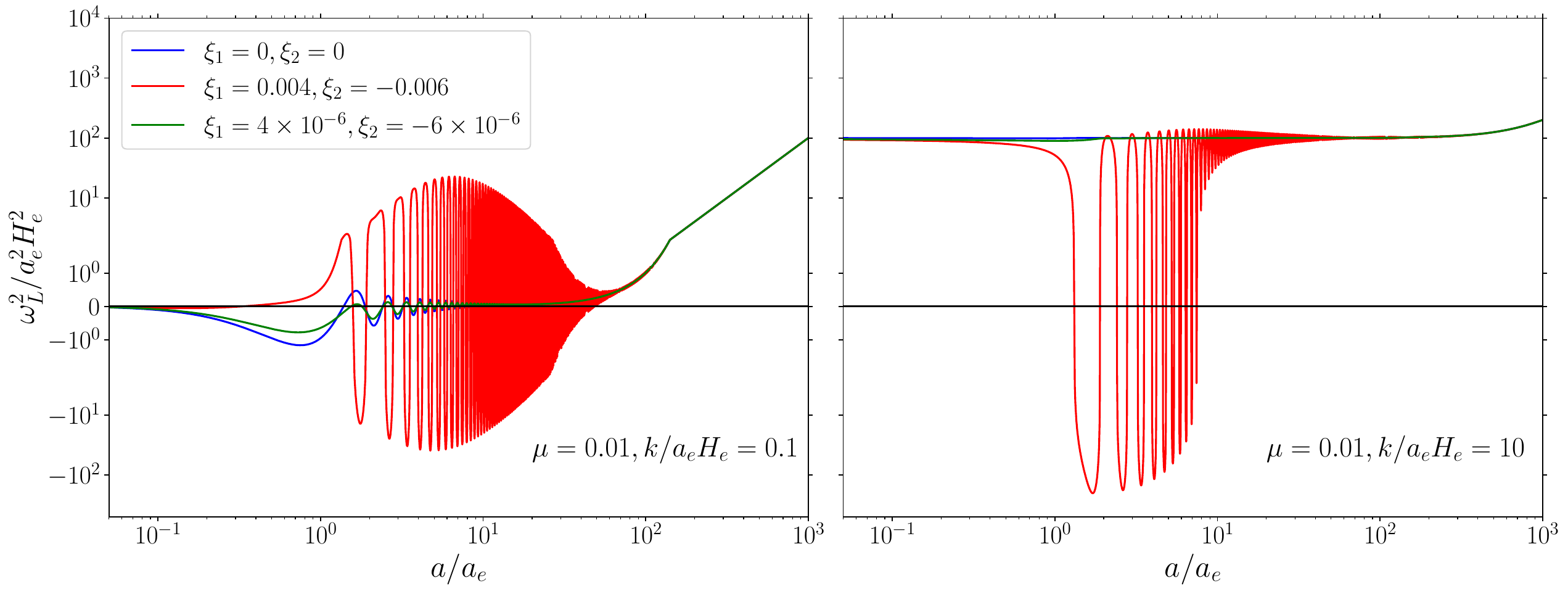}
    \caption{Comparison of $\omega_L^2$ for a mode with $k = 0.1$ vs $k=10$ for three choices of $\{\xi_1,\xi_2\}$. The red curve corresponds to a ``runaway'' mode, which continues to oscillate through zero at arbitrarily large $k$.}
    \label{fig:omega-runaway}
\end{figure}

Let us consider the effects of $c_s^2 < 0$ on $\omega_L^2$. Figure~\ref{fig:omega-runaway}, shows the evolution of $\omega_L^2$ comparing a minimally-coupled model (blue) to a nonminimally-coupled model with $(\xi_1,\xi_2)$ which has $c_s^2< 0$ (red) and a nonminimal model with $(\xi_1,\xi_2)$ where $c_s^2>0$ throughout the evolution (green) for a `low-$k$' ($k/a_eH_e = 0.1$, left) and 'high-$k$' ($k/a_eH_e = 10$, right) mode. 

\begin{figure}
\centering
    \includegraphics[width=0.8\textwidth]{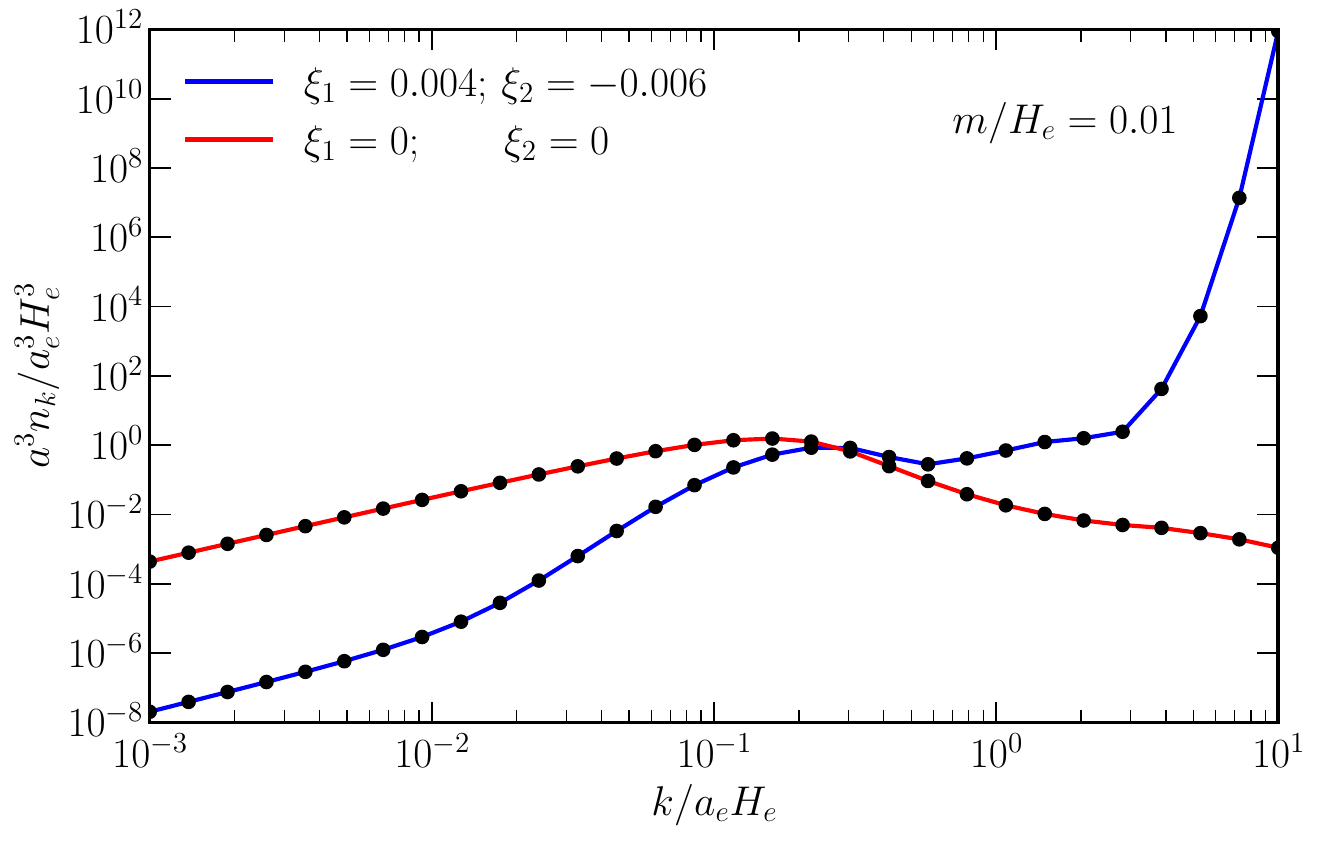}
    \caption{Spectrum of particle production comparing the minimally coupled theory to the nonminimal theory which exhibits a runaway instability.}
    \label{fig:runaway-nk}
\end{figure}

We can see that all three modes are tachyonic at $k=0.1$, oscillating through zero. As mentioned previously, this is not problematic and occurs for even the minimally coupled mode. However, as one goes to larger $k=10$, the minimal and $c_s^2 >0$ modes approach a constant $\omega_L^2 \rightarrow k^2$, but the $c_s^2 < 0$ mode continues to oscillate through zero. One can interpret oscillation through zero as an enhancement of particle production, which thus occurs in this case for arbitrarily large $k$. This is the source the runaway particle production and instability, which we explicitly show an example of in Fig.~\ref{fig:runaway-nk} (from \cite{Capanelli:2024pzd}).

Here we observe that the particle production in the nonminimal theory is actually {\it suppressed} at low $k$, but quickly becomes exponentially enhanced at high $k$. Recall that to determine the comoving number density, one needs to integrate over the $n_k$ spectrum. This is obviously not possible in this case without imposing a cutoff in $k$. In order for the GPP of nonminimally coupled dark photons to be well-behaved and viable as a dark matter production mechanism, one needs to either resolve or avoid this runaway production. The runaway instability found here is similar to a production instability previous found for gravitinos \cite{Kolb:2021xfn,Kolb:2021nob}.

The dark photon runaway instability is discussed in further detail in Ref.\ \cite{Capanelli:2024pzd}, where it is shown that several potential resolutions, including the inclusion of higher-order operators in the Proca EFT, the imposition of a UV cutoff, and the UV completion into a Higgs mechanism, do not 
tame the runaway\footnote{It was recently suggested in \cite{Hell:2024xbv} that the runaway instability can be removed by introducing a disformal transformation such that $g_{\mu\nu} = \eta_{\mu\nu} + h_{\mu\nu} - \beta M_{pl}^{-2}A_\mu A_\nu$. At face value, this solution appears to resolve the runaway, but at the expense of modifying gravity. Furthermore, as was discussed in \cite{Capanelli:2024pzd}, even if one removes the nonminimal couplings responsible for the runaway at tree level, one still expects them to be generated from loops.}. However, we can determine the  $\{\mu, \xi_1, \xi_2\}$ parameter space where the runaway instability does not appear. To avoid catastrophic high-$k$ production, we must require $c_s^2> 0$ throughout inflation and the resuming matter/radiation dominated era. We have already required that $m_t^2 >0$ to prevent ghosts, so we now must also impose that $m_x^2 >0$, leading to the constraint 
\be 
\bar{\mu}_x^2 = \bar{\mu}^2 - \frac{R}{H^2}\left(\xi_1 + \frac{\xi_2}{6}\right) > 0, 
\ee 
corresponding to 
\be 
\bar{\mu}^2 \gtrsim 6\xi_1
\label{eq:muxconstraint}
\ee 
when one considers $-12 < R/H^2 < 6$. Notice that taken along with Eq.~\eqref{eq:munoghosts} this implies that in the massless limit, $\{\xi_1, \xi_2\}$ \textit{must be $\{0,0\}$ to avoid both a ghostly theory and the runaway instability.} 

Fortunately, as one increases the particle mass, there exists an overlapping region in the $\{\xi_1, \xi_2\}$ plane in which both conditions Eqs.~\eqref{eq:munoghosts} and \eqref{eq:muxconstraint} are satisfied. We show below in Fig.\ \ref{fig:noghostnotach} the parameter space for two representative choices of the mass \cite{Capanelli:2024pzd}. We consider $m/H_e=0.1$ (left) and $m/H_e=1$ (right). We see that for any given value of nonzero $\mu$ there is indeed a small region of the $\{\xi_1, \xi_2\}$ plane which avoids both instabilities. For the remainder of the paper, we will focus on this allowed region of parameter space for a given $\mu$. Note that the boundaries of the allowable region are rounded due to the time dependence of $R$ and $H$.

\begin{figure}[htb!]
\centering
  \includegraphics[width=0.49\textwidth]{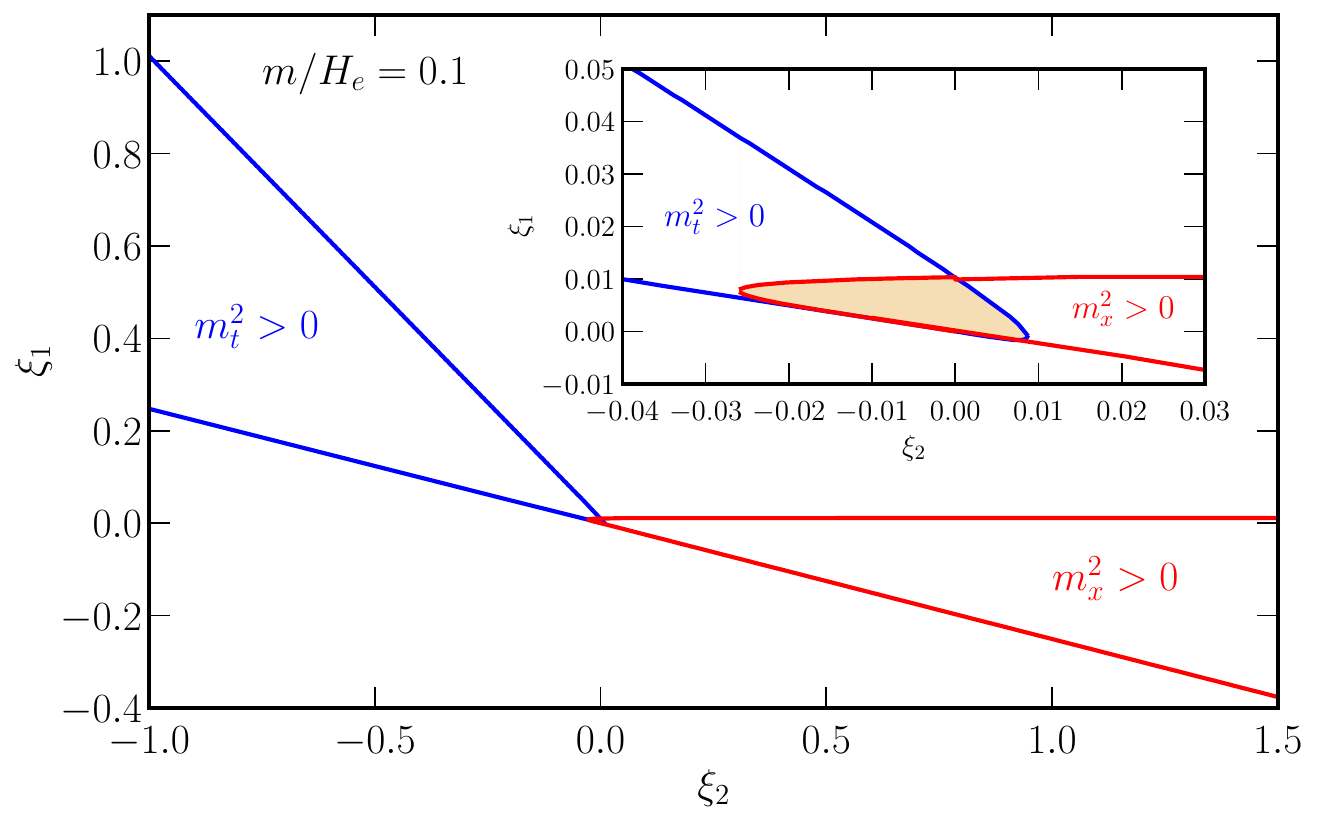} 
        \includegraphics[width=0.49\textwidth]{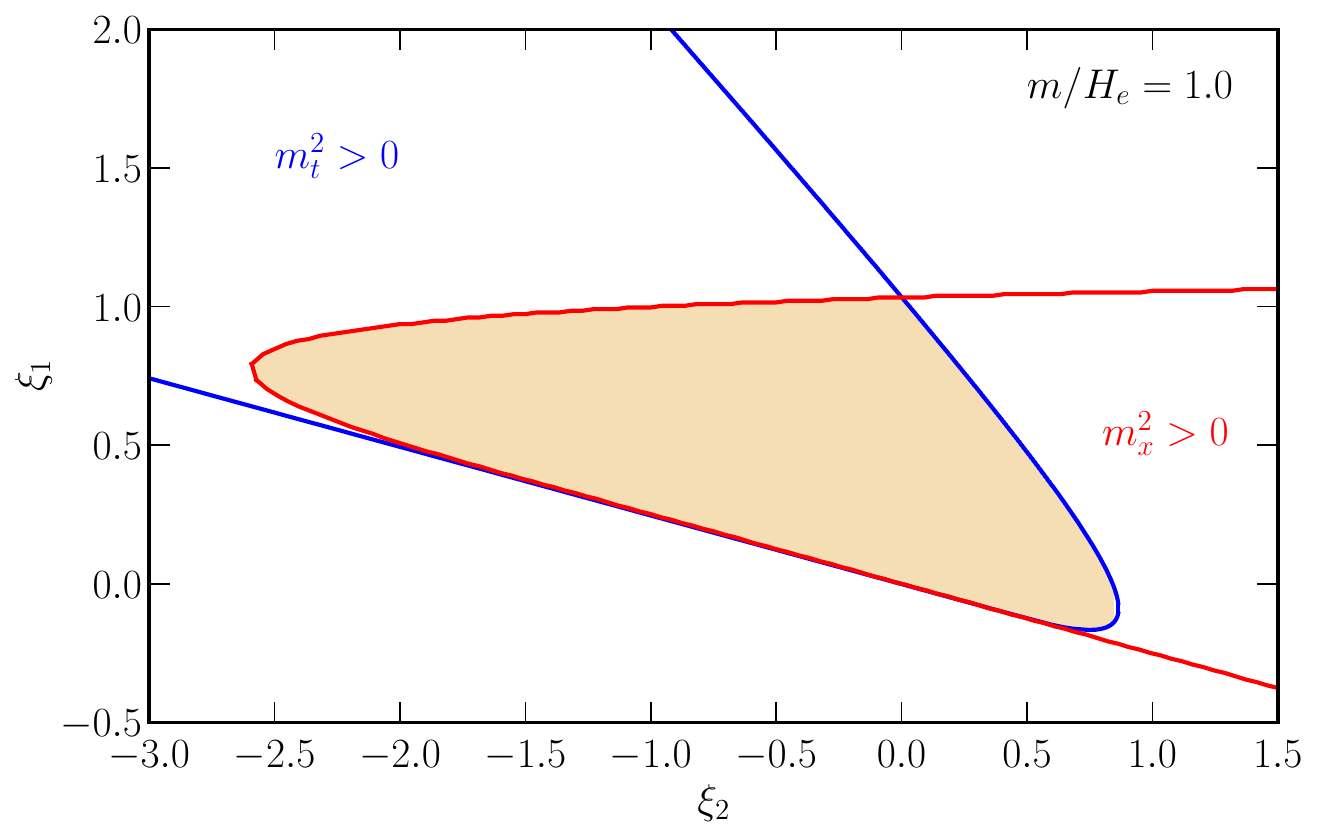}
    \caption{The parameter space in the $\xi_2-\xi_1$ plane for $\mu = 0.1$ (left) and $\mu=1$ (right). The interior of the blue contour has $\mu_t^2>0$, and is therefore ghostless, while the interior of the red contour has $\mu_x^2>0$, and avoids runaway GPP at high $k$. The shaded region is the allowed region to avoid both instabilities.}
    \label{fig:noghostnotach}
\end{figure}

Lastly, note that while the above results are shown for quadratic inflation, the allowed parameter space in $\{\xi_1, \xi_2\}$ is generally \textit{independent of the background inflation model.} Regardless of the specifics of the model, as long as the universe begins in an inflationary de Sitter phase and goes through a kination phase, the same bounds will apply as for the quadratic model above.

\subsection{Superluminal Propagation}

\begin{figure}[htb!]
\centering
    \includegraphics[width=0.85\textwidth]{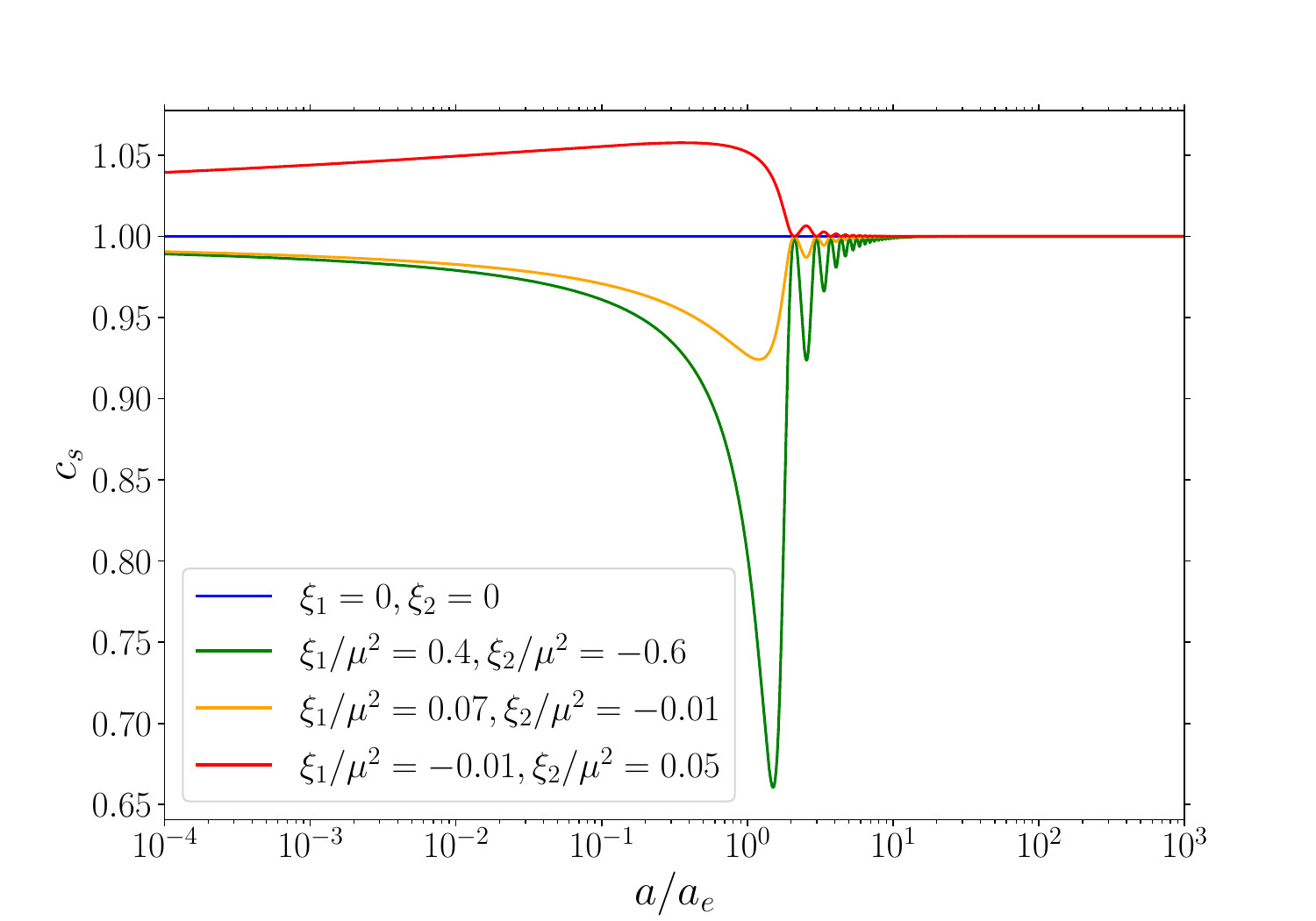}
    \caption{Evolution of the sound speed, $c_s$, for example values of $\{\xi_1, \xi_2\}$ in the ghost-free and runaway-free region. The sound speed begins to oscillate at the end of inflation with amplitude determined by the size of the nonminimal coupling parameters.}
    \label{fig:cs_evol}
\end{figure}

Finally, we note that the sound speed $c_s^2$, Eq.~\eqref{eq:cs2}, can be greater than unity for certain values of $\xi_1$ and $\xi_2$, corresponding to superluminal propagation. To appreciate this, in Fig.\ \ref{fig:cs_evol} we plot the evolution in time of $c_s^2$ for various values of the nonminimal couplings. The red curve exemplifies an example for $c_s^2>1$, corresponding to superluminal propagation. Fortunately, this pathology only occurs for $\{\xi_1 < 0, \xi_2 > 0\}$, and can be avoided by the restriction away from this region, e.g. by the restriction to $\{\xi_1 > 0, \xi_2 < 0\}$.

There are of course theories in which superluminal propagation of perturbations does not lead to causality paradoxes, such as k-essence models \cite{Babichev:2007dw} or noncommutative geometries \cite{Hashimoto:2000ys}. However, there are others in which superluminal propagation is indeed causally problematic (e.g. \cite{Adams:2006sv, DeFelice:2006pg}). A full analysis of the causality structure of the theory is beyond the scope of our current analysis, and so to be conservative, we choose to impose (sub)luminality restrictions on the sound speed. Even with this restriction, the theory admits a broad parameter space for a viable cosmological relic density of dark photons.

\section{GPP of the Nonminimal Proca}
\label{sec:results}
Having enumerated the parameter space where the nonminimally coupled Proca theory is well behaved and instability-free, we now turn to a calculation of the gravitational particle production and the subsequent relic density of dark photon dark matter.

\subsection{Particle Production}
\label{sec:GPPresult-chaotic}

For the numerical results presented in this section, we employ for a background model quadratic inflation with late reheating. For the initial conditions, we take the limit of $\omega_L$ as $a\rightarrow 0$ and $a^2 R \rightarrow 0$ to obtain 
\be 
\omega_L^{2, {\rm init}} = k^2 \frac{m_x^2}{m_t^2} + \frac{(m_t^{2'})^2}{4 m_t^2} - \frac{m_t^{2''}}{2 m_t^2}
\ee 
and define a quantity $k_{\rm eff} = \sqrt{\omega_L^{2, \rm{init}}}$ such that 
\be 
\chi_k^{\rm init} = \frac{1}{\sqrt{2 k_{\rm eff}}} e^{-i k_{\rm eff} \eta}\left( 1 - \frac{i}{k_{\rm eff}\eta}\right),
\ee 
where we have now kept the subleading term in the usual Bunch-Davies initial conditions as well. 

\begin{figure}[htb]
    \centering
    \includegraphics[width=0.9\textwidth]{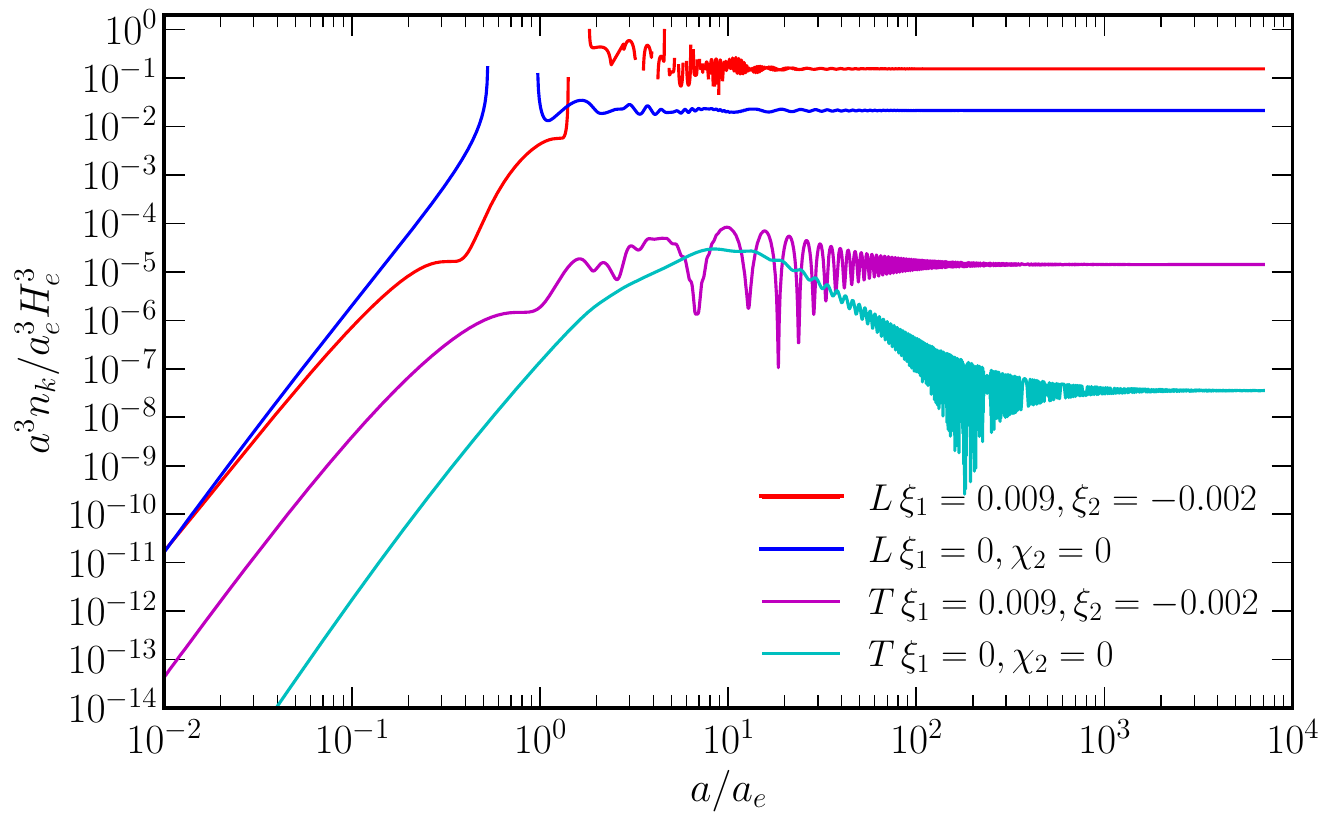}    \caption{
    Evolution of the comoving number density for the specific case of $\mu = 0.1$ and $k/a_eH_e = 1$. We compare the minimal and non-minimally coupled scenarios for particular example values of the nonminimal couplings. In both figures $L$ and $T$ are for the longitudinal and the transverse modes.}
    \label{fig:field-quadratic}
\end{figure}

Recall from the discussion in Sec.~\ref{sec:proca}, the key distinction between the free theory and non-minimally coupled case is the form of the squared frequencies for both the transverse and longitudinal modes arising from their effective masses. In particular, even in the $\{\xi_1, \xi_2\}$ regime which avoids the instabilities discussed in Sec.~\ref{sec:ghosts}, there is still the possibility of a tachyonic instability in $\omega_L^2$ which has the effect of enhancing the total particle production but will remain well behaved at high $k$. We show an example of this in Fig.~\ref{fig:field-quadratic}. We chose a particular mode with $\mu = 0.1$, $k/a_eH_e=1$ and $\{\xi_1 = 0.009, \xi_2 = -0.002\}$.\footnote{Note that the expression for $|\beta_k|^2$ in Eq.\ \ref{eq:betak} has a physical interpretation related to the number density only at late times ($\eta\to\infty$). At earlier times, one may still use Eq.\ \ref{eq:betak} to define $|\beta_k|^2(\eta)$, but when $\omega_k^2\leq0$ the result is imaginary. Regions of $a/a_e$ where $\omega_k^2<0$ are left blank in Fig.~\ref{fig:field-quadratic}.} We show the comoving number density for both the longitudinal and transverse modes. For these particular values, we see that we are in a regime where the particle production is enhanced compared to the minimal theory, but does not lead to runaway production at high $k$. Additionally, while the number density of the transverse mode is enhaced relative to the minimal theory, the particle production is still dominated by the longitudinal mode.

\begin{figure}
    \centering
    \includegraphics[width=0.99\textwidth]{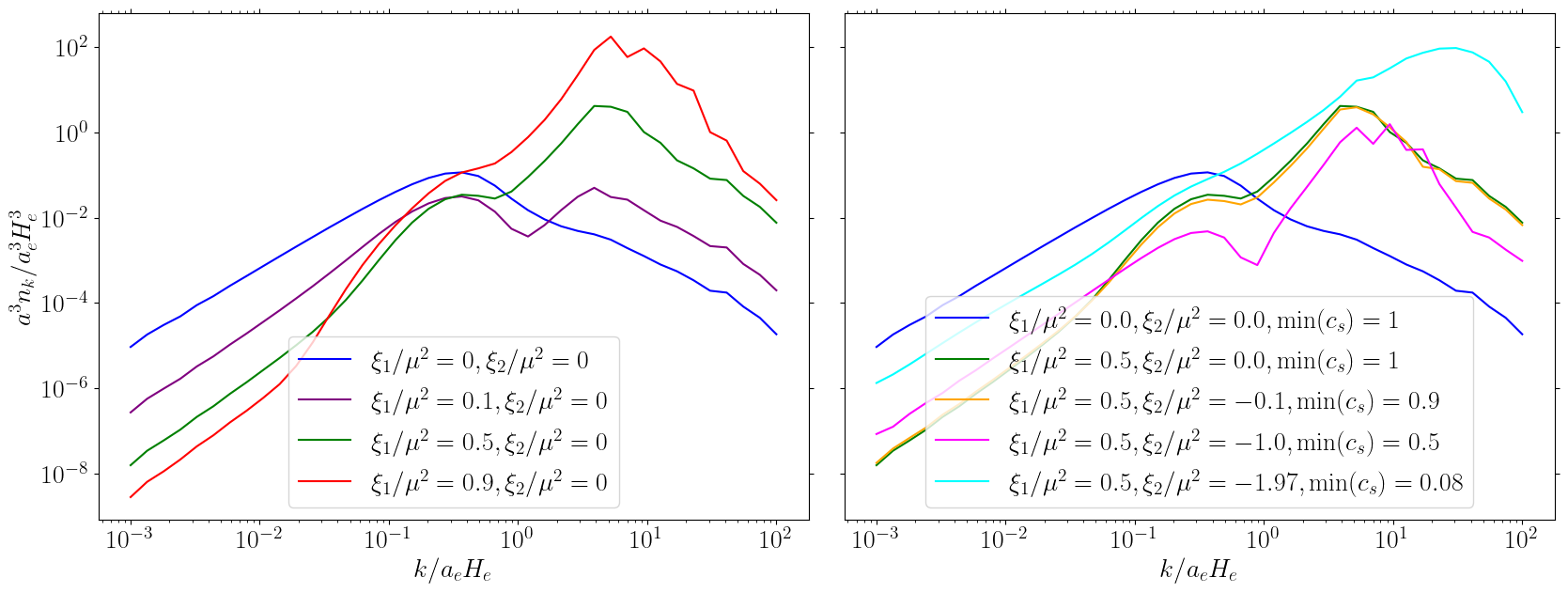}
    \caption{Number density spectrum as a function of $k$ for variations in $\xi_1/\mu^2$ with $\xi_2/\mu^2 = 0$ (left) and fixed $\xi_1/\mu^2$ and nonzero $\xi_2/\mu^2$. We consider $\mu=0.1$ as a representative example. }
    \label{fig:quadratic-spectrum}
\end{figure}

We can also explore how the nonminmal couplings impact the spectrum of $n_k$ as a function of $k$ for a range of masses. In Fig.~\ref{fig:quadratic-spectrum}, we show the comoving number density spectrum, $a^3 n_k/a_e^3 H_e^3$ for a representative example of $\mu=0.1$, comparing the particle production
for the minimal and nonminimal theories.  In this case, we focus solely on the longitudinal modes, as the total number density is dominated by their production. In the left-hand panel we show  $n_k$ for varying $\xi_1$ with $\xi_2 = 0$, and the right-hand panel shows the effect of additionally varying $\xi_2$ for a fixed nonzero value of $\xi_1$. In both cases we compare to the minimal theory with $\{\xi_1 = 0, \xi_2 = 0\}$ (blue curve). On the left, we see that the particle production can be enhanced at high-$k$ for choices of the couplings within the allowable region. In this case, $c_s = 1$ and does not oscillate, and so the enhancement arises from the last three terms in Eq.~\eqref{eq:omegaLfull}. We see that there is a characteristic second peak in the spectrum, near $k/a_eH_e \approx 10$ for the nonminimally coupled spectra, which leads to an enhancement compared to minimal curve (blue). For small enough values of $\xi_1$, as in the purple curve, there is still a characteristic second bump, but the overall production can in fact be suppressed compared to the in the minimal coupling scenario. On the right, we now consider nonzero $\xi_2$ for fixed $\xi_1/\mu^2 = 0.5$. In this case, there is an enhancement contribution from both the last three terms in Eq.~\eqref{eq:omegaLfull} as well as the oscillation of $c_s$. We consider three example values of $\xi_2/\mu^2$, compared to the minimally coupled theory (blue) and the nonminimal theory with $\{\xi_1/\mu^2 =0.5, \xi_2/\mu^2 =0\}$ (green). The yellow curve corresponds to ${\rm min}(c_s) \approx 0.9$, and follows the green curve closely. The magenta curve has ${\rm min}(c_s) \approx 0.5$, but is suppressed compared to the green and yellow because it is in the region of parameter space in which $\xi_2 = -2\xi_1$. In this case, the $R$ dependence in $\mu_t^2$ cancels out and thus the enhancement is lessened. Lastly, the cyan curve corresponds to the very edge of the allowable parameter space, such that ${\rm min}(c_s) \approx 0.08$, which we can see corresponds to the largest amount of enhancement in the produced particle number.

While the above results are for quadratic inflation, the story is qualitatively similar for alternative inflation models. For example, in Appendix ~\ref{sec:GPP-multi}, we show explicitly the GPP for dark photons in two different rapid-turn multi-field inflation models; hyperbolic inflation and monodromy inflation.

\subsection{Parameter Scan Approach: Tracking the Sound Speed}
\label{sec:cs}

The nonminimal Proca theory is characterized by three parameters $\mu$, $\xi_1$, and $\xi_2$. In Sec.~\ref{sec:GPPresult-chaotic}, we considered only a few representative examples for the values of $\{\xi_1, \xi_2\}$. We would like to understand the full parameter space. However, it is computationally expensive to compute the particle production for even a single set of parameters. The latter presents a challenge to fully exploring the $\{ \mu ,\xi_1,\xi_2\}$ parameter space.

\begin{figure}[htb!]
\centering
\includegraphics[width=0.8\textwidth]{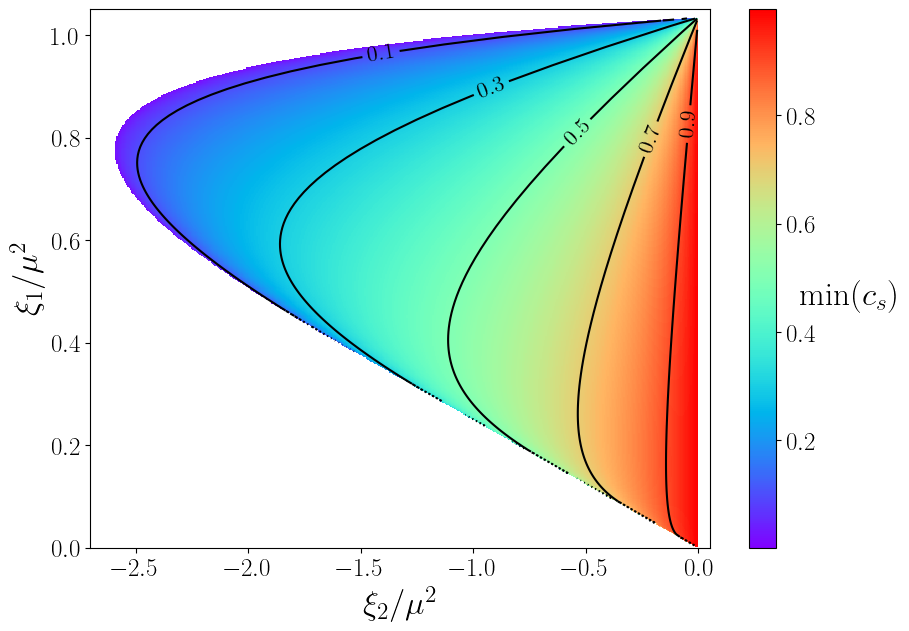}
\caption{ Variation in the minimum oscillation value of $c_s$ over the allowable (ghost-free, non-runaway, subluminal) parameter space of $\xi_{1,2}/\mu^2$ with contours of constant ${\rm min}(c_s)$.  }
\label{fig:cs_full}
\end{figure}

To aid in the parameter scan, and motivated by the sound-speed induced runaway \cite{capanelli2024runaway}, we will sample parameter space according to the behaviour of $c_s ^2$. Namely, we characterize the deviation from $c_s = 1$ by considering the minimum value reached by $c_s$,  corresponding to the first oscillation near $a/a_e = 1$ (see Fig.~\ref{fig:cs_evol}). Since $c_s$ depends on ${\mu,\xi_1,\xi_2}$ only via the combination ${\xi_1/\mu^2,\xi_2/\mu^2}$, this effectively reduces the dimension of the problem from 3 to 2. Figure \ref{fig:cs_full} shows the variation of ${\rm min}(c_s)$ over the allowable (ie., stable) parameter space of ${\xi_{1,2}/\mu^2}$.

\subsection{Contribution to the Present-Day Relic Density}
\label{sec:relic}

Having numerically calculated the comoving number density of dark photons produced during inflation, we would now like to consider how this translates to the relic density today. To obtain a present-day relic density, one has to make assumptions about the time and temperature of post-inflationary reheating, characterized by the parameter $T_{\rh}$. In terms of $\Omega_\chi h^2$, one finds the relic abundance to be \cite{Kolb:2020fwh}
\be \label{eq:omegachi}
\frac{\Omega_\chi h^2}{0.12} = \frac{m_\chi}{H_e}\left(\frac{H_e}{10^{12}\text{GeV}}\right)^2 \left(\frac{T_\rh}{10^9 \text{GeV}}\right)\frac{na^3}{10^{-5}} \ .
\ee 
For simplicity we will consider a fixed $T_\rh = 10^6$ GeV and $H_e=10^{12}$ GeV. Changing the values of these parameters simply leads to an overall scaling in the value of $\Omega_\chi h^2/0.12$ as in Eq.\ \ref{eq:omegachi}. Recall that to obtain the correct present-day abundance, one requires that $\Omega_\chi h^2/0.12 \approx 1$.

To explore the the dependence of the relic density on  the parameters $\xi_{1,2}$ and $\mu^2$ we follow the approach discussed in Sec.~\ref{sec:cs} and compute the present-day relic density along contours of constant ${\rm min}(c_s)$ as shown in Fig.~\ref{fig:cs_full}. Since the the particle production is  dominated by the longitudinal mode (see Fig.~\ref{fig:field-quadratic}), we neglect any contribution from the transverse modes.

\begin{figure}[htb!]
    \centering
   \includegraphics[scale=0.35]{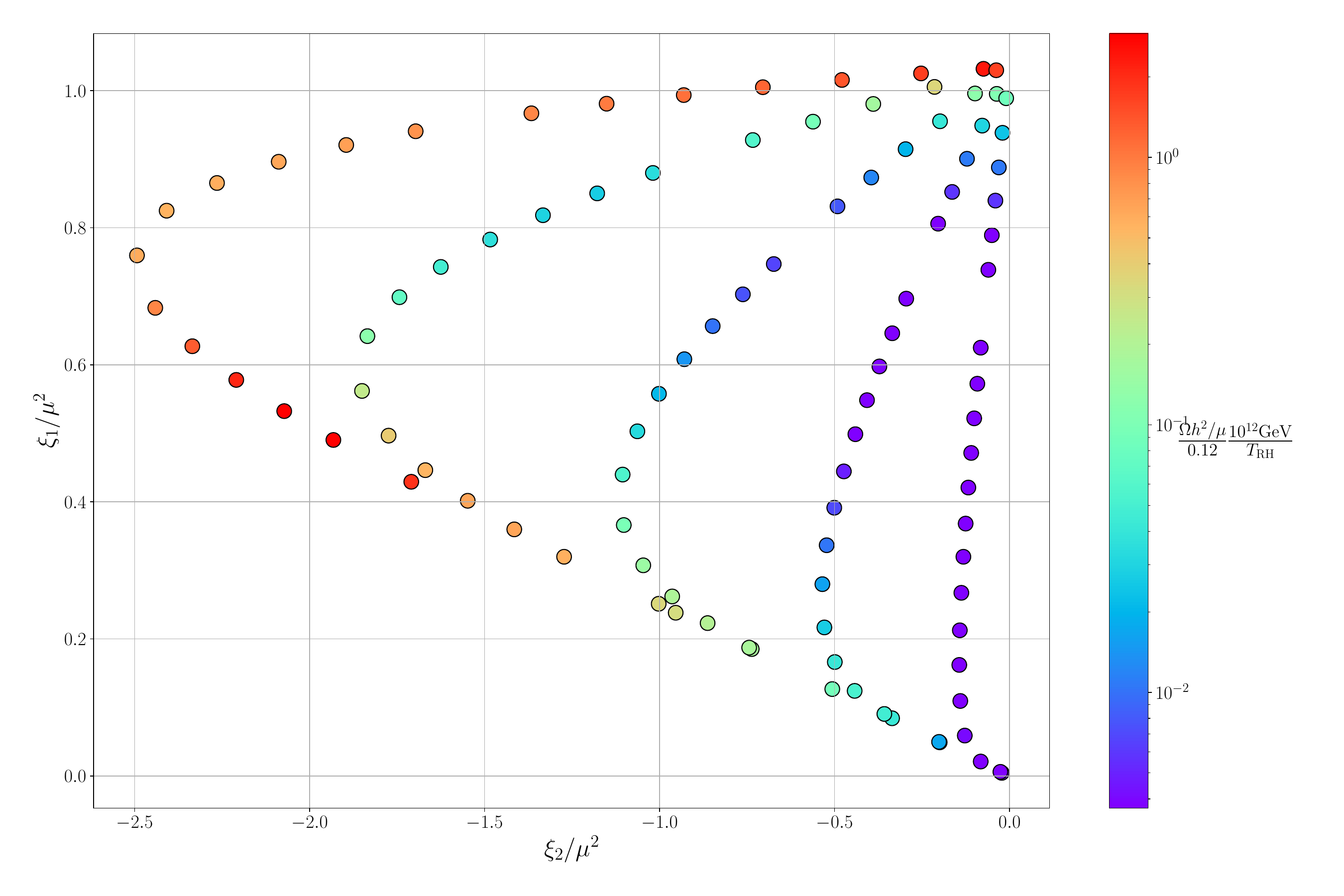}

    \caption{Relic density $\Omega h^2$ in the $\{\xi_1,\xi_2\}$ plane for $\mu=0.1$.}
    \label{fig:relic-plots}
\end{figure}

Results for $\Omega_\chi h^2$ are shown in Fig.~\ref{fig:relic-plots}. Here we fix $\mu=0.1$ and sample in $\xi_{1,2}$ according along contours of fixed min$(c_s)$. The results are shown in terms of the quantity $(\Omega_\chi h^2/\mu)/0.12 \times (10^{12} {\rm GeV})/T_{\rm RH}$.

We can see that for these choices of $\mu, H_e$, we obtain a range of $\mathcal{O}(10^-3) \lesssim \Omega_\chi h^2/0.12 \lesssim \mathcal{O}(1)$. Recall that the present-day relic density corresponds to $\Omega_\chi h^2/0.12 \approx 1$. One can easily change $H_e$ or $T_\rh$ to obtain the correct amount of DM. 

\begin{figure}
    \centering
    \includegraphics[scale=0.35]{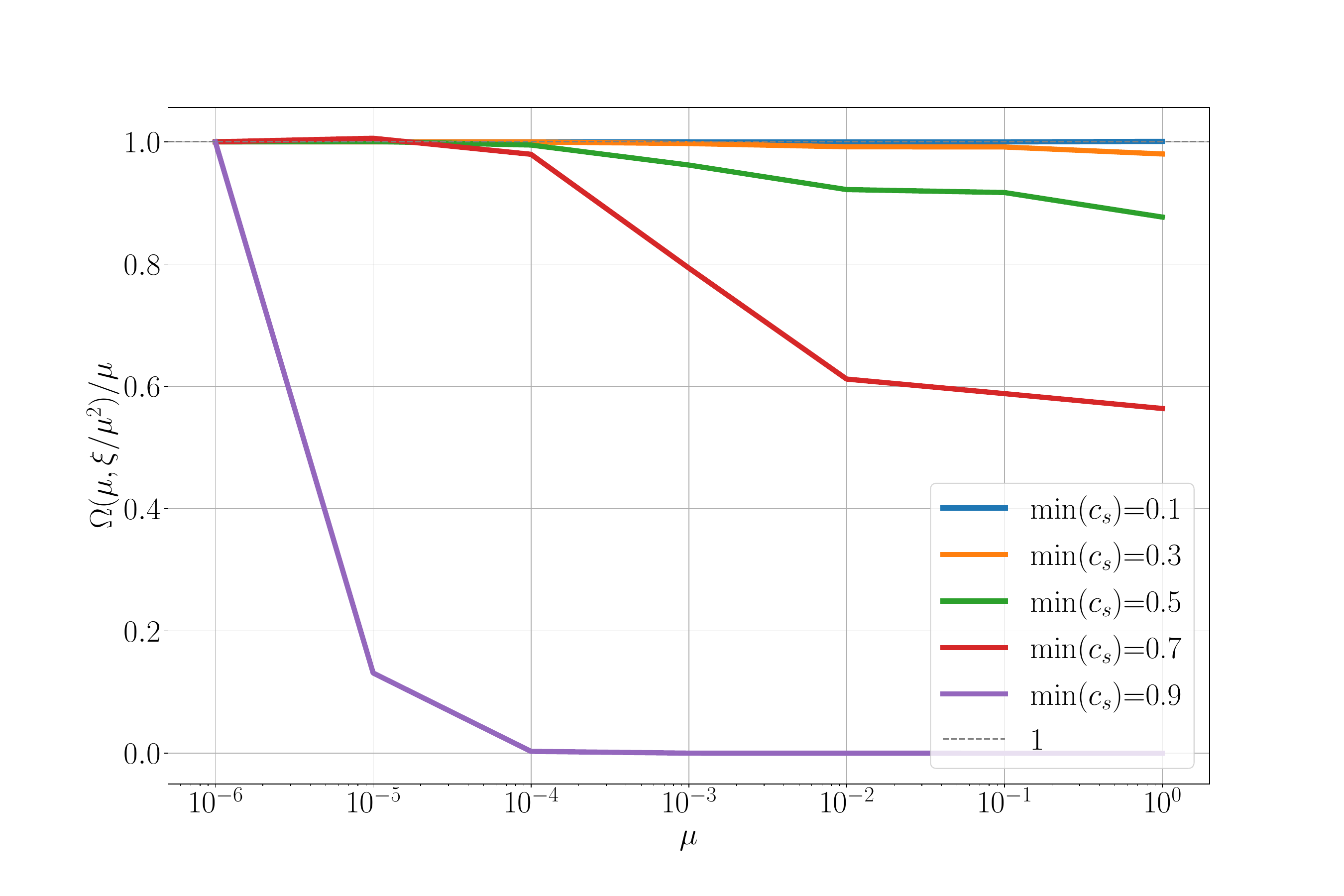}
    \caption{$\Omega_\chi h^2$ as a function of particle mass $\mu$ for fixed $\xi_{1,2}/\mu^2$ (and hence fixed ${\rm min}(c_s)$). }
    \label{fig:mspec}
\end{figure}

Furthermore, for parameters with min$(c_s) \ll1$, the dependence of the relic density on the mass $\mu$ is approximately linear, reflecting the overall prefactor of $m_\chi/H_e$ in Eq.~\eqref{eq:omegachi}. This indicates that the produced comoving particle number, $na^3$ is \textit{independent of the mass}. The latter manifests the $\mu$-rescaling invariance of $c_s ^2$ and the fact that the particle production is driven by the reduction of $c_s$ in this regime. This can be appreciated Fig.~\ref{fig:mspec}, which shows $\Omega_\chi h^2/0.12$ as a function of the particle mass at fixed values of $\{\xi_1/\mu^2, \xi_2/\mu^2\}$

From Fig.~\ref{fig:relic-plots} we see that, throughout the parameter space, the relic density tracks the increasing contours of ${\rm min}(c_s)$, reaching a maximum towards the edge of the parameter space where one approaches the runaway region. As a result, it is possible that one can obtain arbitrarily light dark photons in a wider range of parameter space than is possible in the minimally coupled scenario. In particular, we expect the possible mass range to be increased for a wider range of $H_e$ and $T_{\rh}$ values. We defer a full analysis of the implications of this scenario of ultralight dark matter to future work.

Finally, in Fig.~\ref{fig:mspec}, we show the relic density as a function of the particle mass for the various ${\rm min}(c_s)$ contours. We see that for the large deviations in $c_s$, $\Omega_\chi h^2/0.12$ is independent of the mass for $\mu \lesssim 1$, then begins decreasing as one goes to higher masses. This matches our intuition from GPP in the minimally coupled theory \cite{Kolb:2020fwh}.

\section{Discussion and Conclusions}
\label{sec:conclusions}
In this work we have extended previous analysis of gravitational particle production of spin-1 dark matter to include non-minimal couplings to gravity. We have first explored the viable parameter space of the theory to avoid both ghosts and a high-$k$ tachyonic instability which leads to runaway particle production. Requiring that any healthy theory of nonminimally coupled dark photons avoid both of these instabilities sets mass-dependent constraints on the value of the coupling parameters $\{\xi_1, \xi_2\}$, and we further restrict to modes which have a subluminal sound speed. We then showed how the addition of allowed nonminimal couplings in the theory impacts GPP of the dark photons. We find that there is a region of $\{\xi_1, \xi_2\}$ which is free of the above instabilities but that still leads to an overall enhancement of the particle number. As a result, we find that GPP can be responsible for production of dark photon dark matter, widening the parameter space for a range of particle masses, inflationary energy scales, and reheating models. Lastly, we comment that the results presented here are largely independent of the background inflation model one chooses and also hold for a wider class of rapid-turn multi-field inflation models.

It is further interesting to note, that as was pointed out in \cite{Capanelli:2023uwv}, one can obtain the dark matter relic density for a class of Kalb-Ramond-like-particle (KRLP) dark matter with an identical procedure to the spin-1 scenario, with and without nonminimal couplings. Thus, the results that we have presented here are also applicable to GPP of KRLP dark matter. 

There are many interesting paths forward. For example, there is much to be done in the realm of particle phenomenology of spin-1 dark matter, including that which is produced gravitationally, in particular dark matter direct detection, which is sensitive to the velocity distribution of dark matter and hence primordial spectrum of particles $n_k$. Complementary to this is {\it directional} direct detection, which is sensitive to the spin of the dark matter particle \cite{Jenks:2022wtj,Catena:2018uae,Catena:2017wzu}. On the astrophyics and cosmology side, it will be interesting to study the implications of nonminimal gravitational couplings, in particular in the case of {\it ultralight} vector dark matter. Spin-1 dark matter production during inflation may also be probed by hot spots in the cosmic microwave background, analogous to that in the case of scalar dark matter production \cite{Maldacena:2015bha,Fialkov:2009xm,Kim:2021ida,Kim:2023wuk} (for recent constraints see \cite{Philcox:2024jpd}).

We leave these and other interesting directions to future work.

\acknowledgments
The authors thank Anamaria Hell, Bohdan Grzadkowski, Andrew Long, and Anna Socha for helpful discussions and correspondence.  C.C.\ is supported by a fellowship from the Trottier Space Institute at McGill via an endowment from the Trottier Family Foundation, and by the Arthur B. McDonald Institute via the Canada First Research Excellence Fund (CFREF).  L.J.\ is supported by the Kavli Institute for Cosmological Physics at the University of Chicago. The work of E.W.K. was supported in part by the US Department of Energy contract DE-FG02-13ER41958 and the Kavli Institute for Cosmological Physics.  E.M.\ is supported in part by a Discovery Grant from the Natural Sciences and Engineering Research Council of Canada and by a New Investigator Operating Grant from Research Manitoba.

\appendix
\section{GPP of Dark Photons in Rapid-Turn Multi-Field Inflation}
\label{sec:GPP-multi}
In this Appendix we give an overview of GPP of nonminimally coupled dark photons in an additional class of multi-field, rapid-turn inflation models. These models have more complicated dynamics than the single field $m^2\varphi^2$ model consdidered above, but we will see that GPP results are qualitatively similar. Such models have gained interest recently due to their consistency with CMB results as well as the de Sitter swampland conjecture. Following the discussion in \cite{Kolb:2022eyn}, we consider two realizations; hyperbolic inflation and monodromy inflation. We provide a brief overview of each of these models below in Sec.~\ref{sec:models} and then numerical results for the GPP in Sec.~\ref{sec:GPPresults-multi}.

\subsection{Rapid-Turn Multi-Field Inflation}
\label{sec:models}
Here we provide an overview of both hyperbolic inflation and monodromy inflation. Further details of the models and their motivations can be found in Ref.\ \cite{Kolb:2022eyn}.

\begin{enumerate}\item \textbf{Hyperbolic Inflation}

We first consider a model of hyperbolic inflation (`hyperinflation'), which is characterized by having a hyperbolic field-space manifold, see e.g., \cite{Brown:2017osf,Bjorkmo:2019aev,Christodoulidis:2018qdw,Garcia-Saenz:2018ifx}. In this case, there are two fields: a radial field, $\phi$, and an angular field, $\theta$, and the dynamics are characterized by the action \cite{Brown:2017osf}
\be 
S = \int d^4 x \sqrt{-g}\left[\frac{1}{2}G_{IJ}(\phi)\partial_\mu \varphi^I\partial^\mu\varphi^J - V(\phi)\right],
\ee 
where $I,J$ stands for $\phi,\theta$ and the field space metric is 
\be 
G_{IJ}(\phi) = \begin{pmatrix} 1  & 0 \\
0 & L^2\sinh^2(\phi/L)\end{pmatrix}, 
\ee 
with $L$ defined in terms of the scalar curvature $R_{\text{field-space}}  = -2/L^2$. For the potential we consider a representative toy model chosen for simplicity: 
\be 
V(\phi) = V_0\tanh^2\left(\frac{\phi-v}{f}\right) e^{\lambda\phi/M_\pl}.
\ee 

\item \textbf{Monodromy Inflation:}

As a second example of multi-field rapid-turn inflation, we additionally consider monodromy inflation \cite{Achucarro:2019pux, Achucarro:2012yr}. In this scenario, the model again has a radial field, $\phi$ and an angular field, $\theta$, but the field-space is now flat. The action is given by 
\be 
S = \int d^4 x \sqrt{-g} \left[\frac{1}{2}(\partial\phi)^2 + \frac{1}{2}\phi^2(\partial\theta)^2 - V(\theta,\phi)\right], 
\ee 
As in Ref.\ \cite{Kolb:2022eyn} we consider a potential of the form 
\be 
V(\theta,\phi) = V_0 e^{-\alpha(\theta - \theta_i)/V_0}\tanh^2(\theta/f) + \frac{1}{2}m^2(\phi - \phi_0)^2.
\ee 
\end{enumerate}

\subsection{GPP Results: Rapid-Turn Multi-Field Inflation}
\label{sec:GPPresults-multi}
We now turn to the GPP dark photons in the two rapid-turn multi-field inflation models discussed above. The parameters for hyperbolic inflation are chosen to be \cite{Kolb:2022eyn}
\begin{align*}
V_0 & = 2.6\times10^{-36}\MPl^4 \\
L & = 0.0054 \MPl \\
\lambda & =2 \\
v & = 10^{-2} \MPl \\
f & = 10^{-2} \MPl
\end{align*}
while for the monodromy model the parameters are takes to be
\begin{align*}
    V_0 & = 7.49\times 10^{-10}\MPl^4\\
    \alpha & = -5.05\times10^{-13}\MPl^4 \\
    f & = 10 \\
    \phi_0 & = 6.80\times 10^{-4}\MPl \\
    m & = 4.00\times10^{-4}\MPl \\
    \theta_i & = 2000 \ .
\end{align*}
These parameters were chosen to be consistent with CMB observations \cite{Planck:2015sxf, Planck:2018jri}.

In this Appendix, we consider the GPP of dark photons \textit{minimally} coupled in the models described above, taking $\xi_1 = \xi_2 = 0$. An analysis of which has not yet appeared in the literature. We anticipate that adding $\xi_{1,2} \neq 0$ will yield similar results as in the quadratic model previously discussed, but we defer a full analysis to future work. 

\begin{figure}[htb!]
\centering 
    \includegraphics[width=0.7\textwidth]{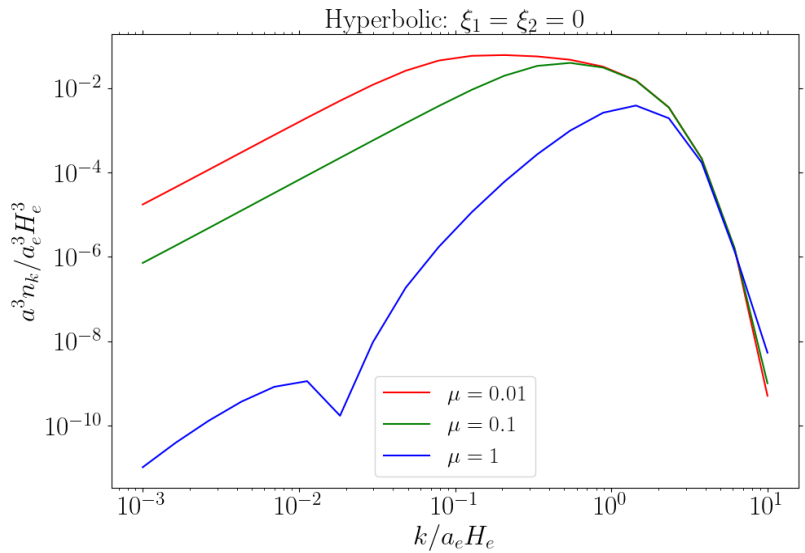}
    \label{fig:hyper-min-spectrum}
    \caption{Comoving number density spectrum of the longitudinal modes as a function of $k$ in the minimally coupled theory assuming hyperbolic inflation for $\mu = 0.01, 0.1, 1.0$. }
\end{figure}

\begin{figure}[htb!]
\centering 
    \includegraphics[width=0.7\textwidth]{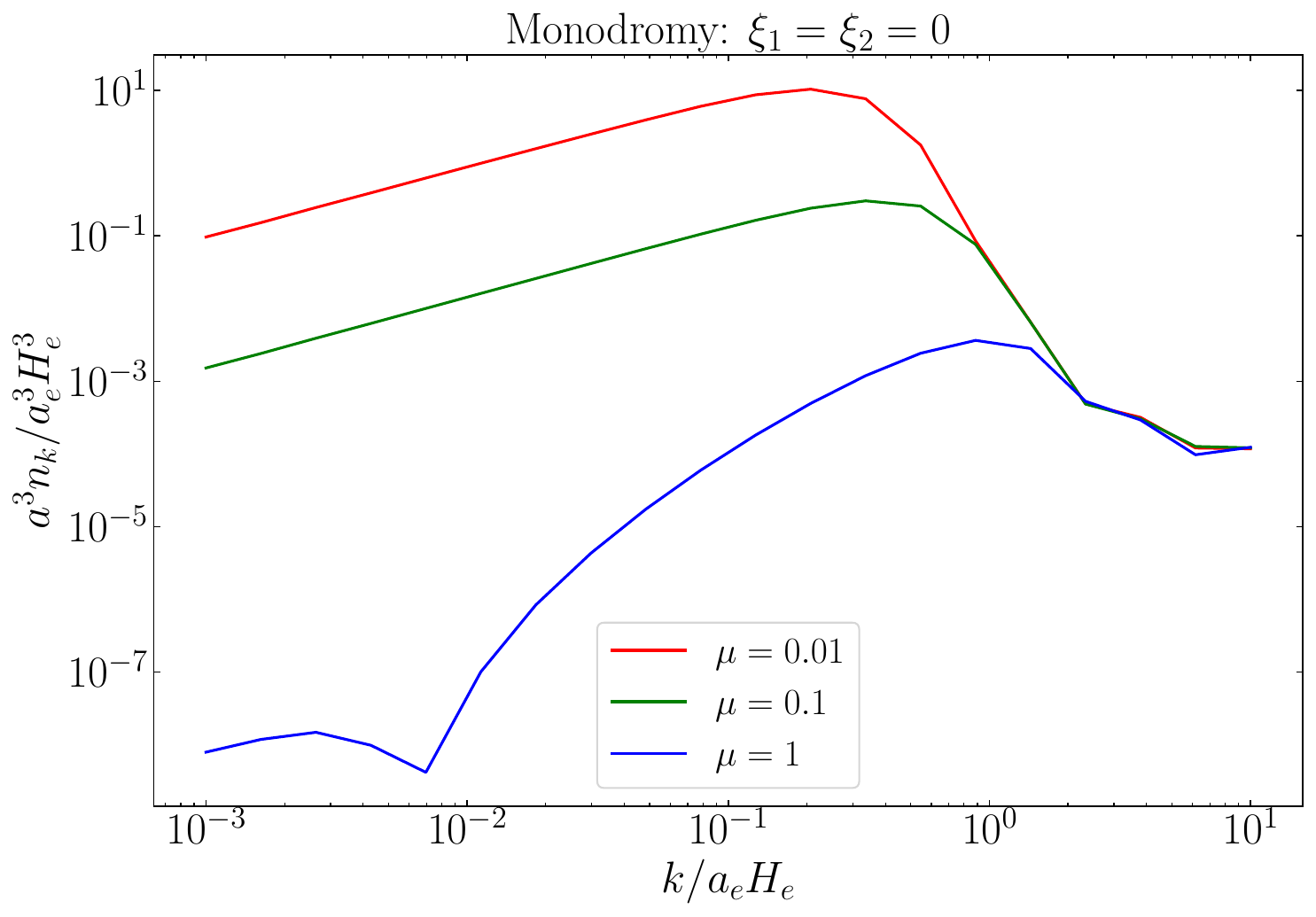}
    \label{fig:mono-min-spectrum}
    \caption{Comoving number density spectrum of the longitudinal modes as a function of $k$ in the minimally coupled theory assuming monodromy inflation for $\mu = 0.01, 0.1, 1.0$. }
\end{figure}

As before, we numerically solve for the spectrum of particle production as shown in Figures~\ref{fig:hyper-min-spectrum} and~\ref{fig:mono-min-spectrum} for hyperbolic and monodromy inflation, respectively. We consider representative masses $\mu = 0.01, 0.1, 1$ and see that the shape of the spectra differ between the two models and the peak value is several orders of magnitude higher in monodromy inflation than for the hyperbolic model. This is in agreement with the results of \cite{Kolb:2022eyn}, which found that while monodromy inflation could be a viable model to obtain the relic density of scalar dark matter via GPP, the particle density produced in hyperbolic inflation was too low to explain the present-day dark matter abundance.

Given that we found the impact of adding nonminimal couplings is to enhance the particle production in the quadratic inflation model, it is reasonable to expect that the same would be the case for both of the multifield models. It is likely that enhancement of the particle density produced in monodromy inflation would widen the range of viable parameter space of e.g. $H_e$ and $T_{\rh}$, as we found for the quadratic model. Similarly, it is possible that an enhancement due to nonminimal couplings could render hyperbolic inflation able to produce the dark matter relic density via GPP. We save a careful analysis of both of these models with the inclusion of nonminimal couplings for future work.


\section{Early Reheating}
\label{sec:early}
How might early reheating effect the runaway instability?  Answer: It doesn't matter much.  In this appendix we justify that statement.

Reheating can be modelled by including a decay width, $\Gamma_\varphi$, for the inflaton field, $\varphi$, into the equation of motion for the inflaton field, and the evolution of the matter and radiation energy densities $\rho_\varphi$ and $\rho_R$ (for more details, see, e.g., Ref.\ \cite{Giudice_2001}):
\begin{align}
   &  \ddot{\varphi} + 3H\dot{\varphi} + dV(\varphi)/dt + \Gamma_\varphi\dot{\varphi} = 0 \ , \\
   & \dot{\rho}_\varphi + 3H\rho_\varphi + \Gamma_\varphi\rho_\varphi = 0 \ , \\
   & \dot{\rho}_R + 4H\rho_R - \Gamma_\varphi\rho_\varphi = 0 \ .
\end{align}
The initial condition deep in the inflationare era is $\rho_R=0$.  

After inflation $\rho_R$ grows to become equal to the $\rho_\varphi$, and thereafter dominates $\rho_\varphi$.  We refer to the epoch when $\rho_R=\rho_\varphi$ as ``reheating,'' with the value of the scale factor at the epoch defined as $a_\mathrm{RH}$. The value of $a_\mathrm{RH}/a_e$ is determined by $\Gamma_\varphi$.  We consider two representative values of $\Gamma_\varphi$: $0.1m_\varphi$ and $0.01m_\varphi$.  The first choice results in $a_\mathrm{RH}/a_e=3.34$ while the second choice results in $a_\mathrm{RH}/a_e=15.52$.  (As mentioned in Sec.~\ref{sec:ghosts}, for late reheating, reheating occurs at a late enough time that during the relevant timescales for particle production the universe is still in a matter-dominated phase.)

\begin{figure}[htb!]
    \includegraphics[width=0.95\textwidth]{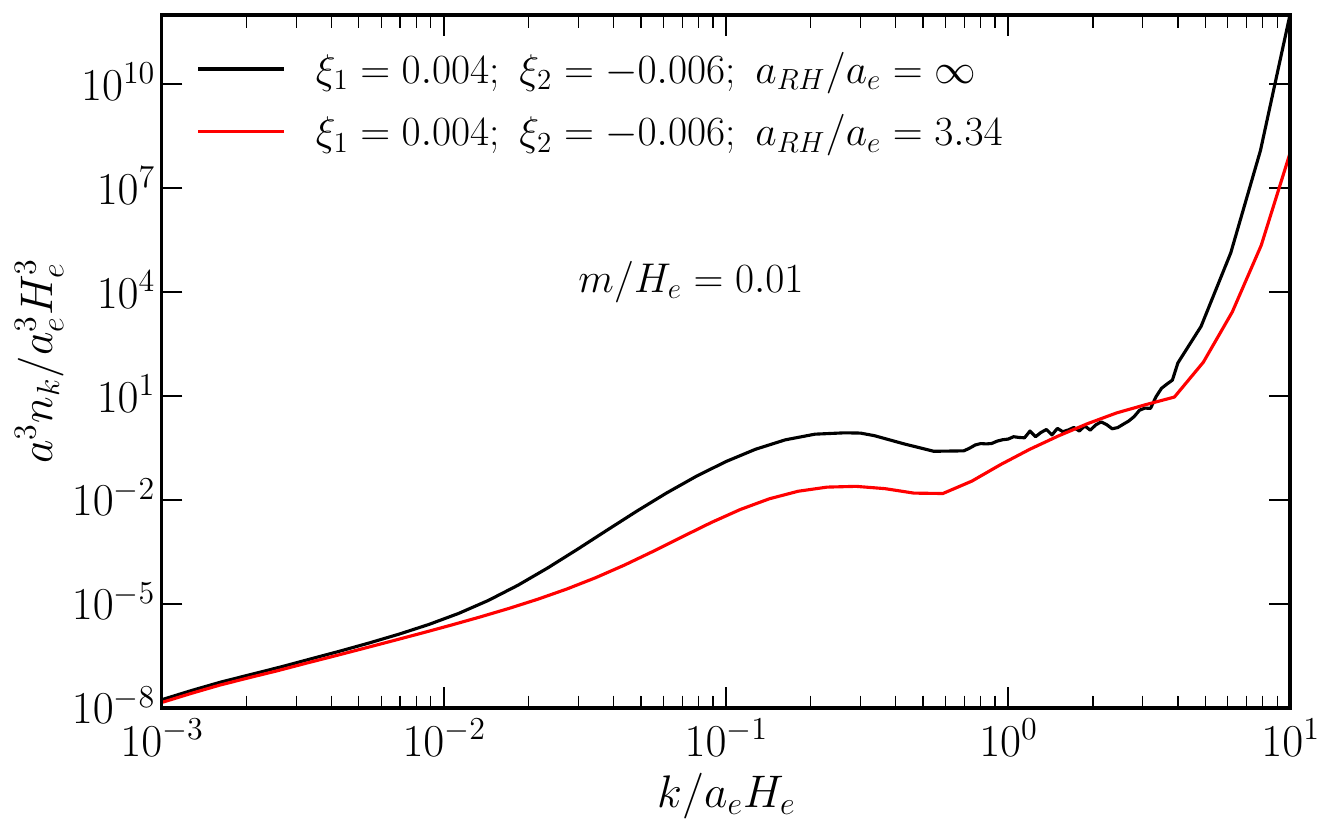}
    \caption{The effect of early reheating on the spectrum of GPP for $m/H_e=0.01$. The festures in $n_k$ for late reheating in the region $1 \lesssim k/a_eH_e \lesssim 3$ are due to quantum interference in gravitational particle production \cite{Basso_2022}.}
    \label{fig:early}
\end{figure}

From Fig.\ \ref{fig:early} we see that early reheating lessens the high-$k$ instability but does not remove it.

\bibliographystyle{JHEP} 
\bibliography{ref}

\end{document}